\documentclass[twocolumn,showpacs,amsmath,amssymb]{revtex4}
\usepackage{graphicx}

\begin{document}
\def\d{\partial}
\def\dx{\partial x}
\def\df{\partial f}
\def\eqn{\begin{equation}}
\def\eqns{\begin{eqnarray}}
\def\endeqn{\end{equation}}
\def\endeqns{\end{eqnarray}}

\title{Long Tailed Maps as a Representation of Mixed Mode
Oscillatory Systems}

\author{Rajesh Raghavan$^2$ and G. Ananthakrishna$^{1,2}$\\
$^1$ Materials Research Centre\\
Indian Institute of Science\\
Bangalore  560012, India\\
\vspace{0.3cm}
{\small and}\\
$^2$ Center for Condensed Matter Theory\\
Indian Institute of Science\\
Bangalore 560012, India
}

\begin{abstract}
Mixed mode oscillatory (MMO) systems are known to exhibit some generic features such as the 
reversal of period doubling sequences and crossover to period adding 
sequences as  bifurcation parameters are varied. In addition, they exhibit a nearly one dimensional unimodal Poincare map with a longtail.  We recover  these common features  
from a general class of two parameter family of one dimensional maps with a unique critical point that satisfy  a few general constraints that determine  the nature of the map. 
We derive scaling laws that determine the parameter widths of the dominant 
windows of periodic orbits sandwiched between two successive 
states  of $RL^k$ sequence.  An example of a two parameter map with a unique
critical point is introduced  to  verify the analytical
results.
\end{abstract}


\maketitle

\section{Introduction}
Dynamical systems with disparate time scales
for the participating modes often exhibit periodic states
characterised by a combination of relatively large amplitude and
nearly harmonic small  amplitude oscillations. Such periodic
states, called the mixed mode oscillations (MMOs),  
and the associated complex bifurcation
sequences have been observed in models and experiments in the area
of chemical kinetics \cite{gyorgi,barkley,petrov92,koper95,bz},
electrochemical reactions  \cite{alba89,koper92,krisher93},
biological systems \cite{chay}, and in many physical systems
\cite{braun,tamosi89,raj00}. These  MMO systems   typically
exhibit the following features - (a) period doubling (PD)
sequences and their reversal in multi-parameter space with
respect to a  primary bifurcation parameter keeping other
parameters fixed, and (b) crossover to bifurcation
sequences of alternate periodic-chaotic windows  when one of
the secondary bifurcation parameters is varied wherein dominant
windows of periodicity 
increase in an arithmetical order which we refer to  as period
adding (PA) sequences
\cite{petrov92,koper95,alba89,koper92,krisher93,raj00,hauser96}.
\\

\begin{figure}
\begin{center}
\includegraphics[width=8.0cm]{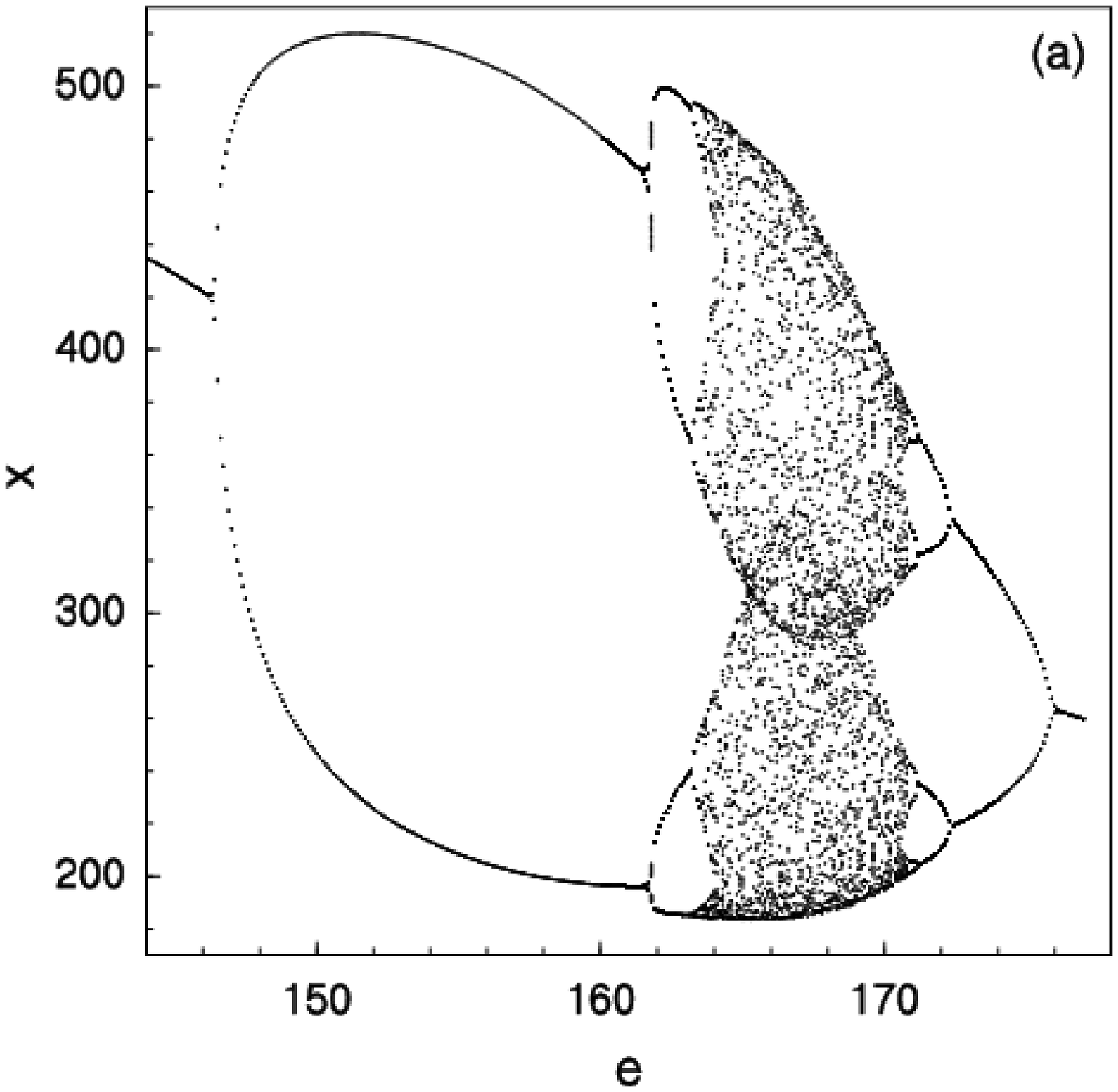}
\includegraphics[width=8.0cm]{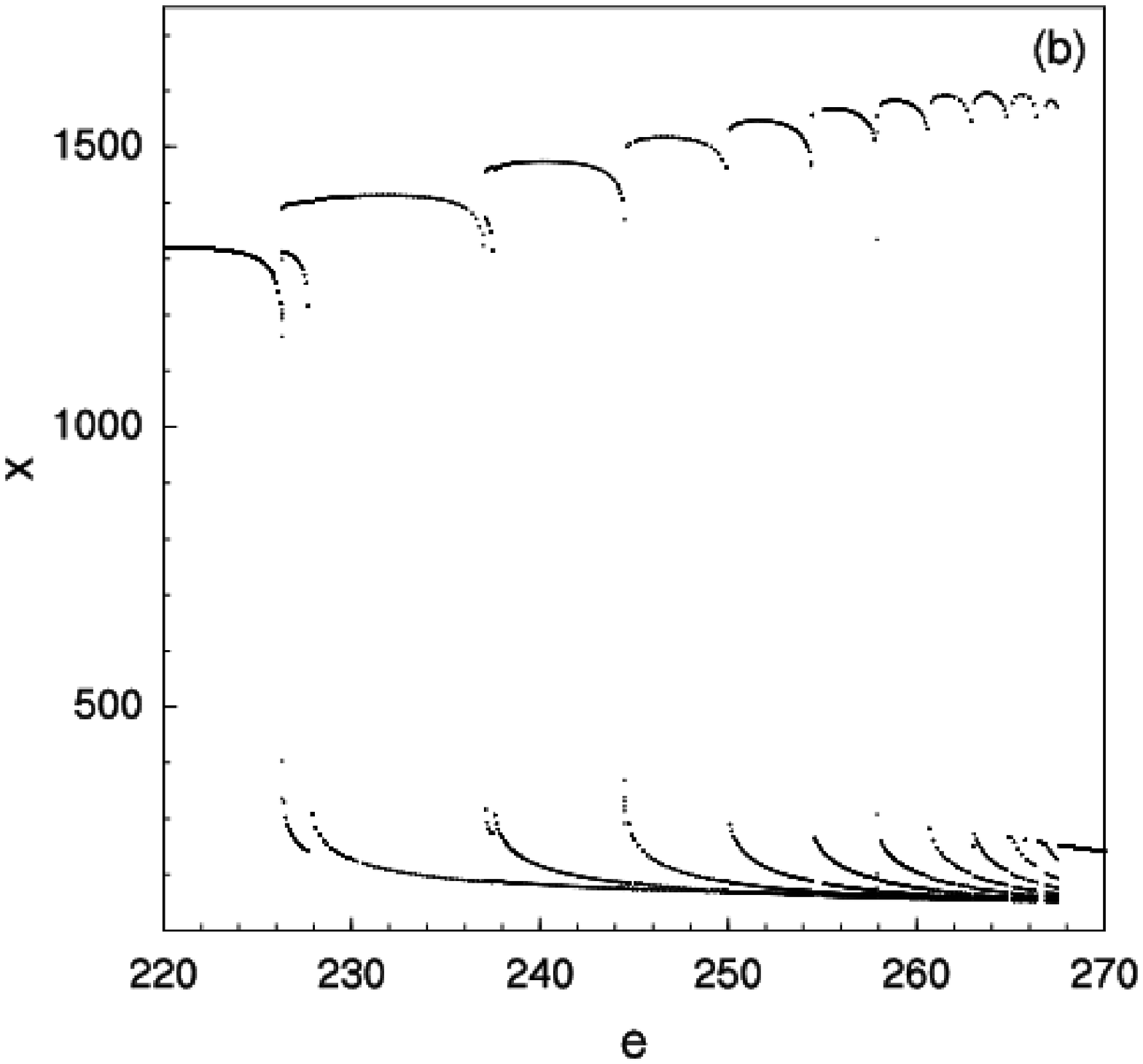}
\caption{ Bifurcation diagrams for the AK model for a plastic
instability with primary bifurcation parameter, $e$ and 
secondary bifurcation parameter, $m$. 
(a) Cascading period doubling bifurcations and their
reversals forming a  bubble structure for  $m = 2.16$, and (b)
period adding sequences for $m=1.2$. Chaotic regions exists in
a vanishingly small parameter regions sandwiched between
successive $1^s$ periodic states. }  \label{fig1}
\end{center}
\end{figure}

\begin{figure}
\begin{center}
\includegraphics[width=8.0cm]{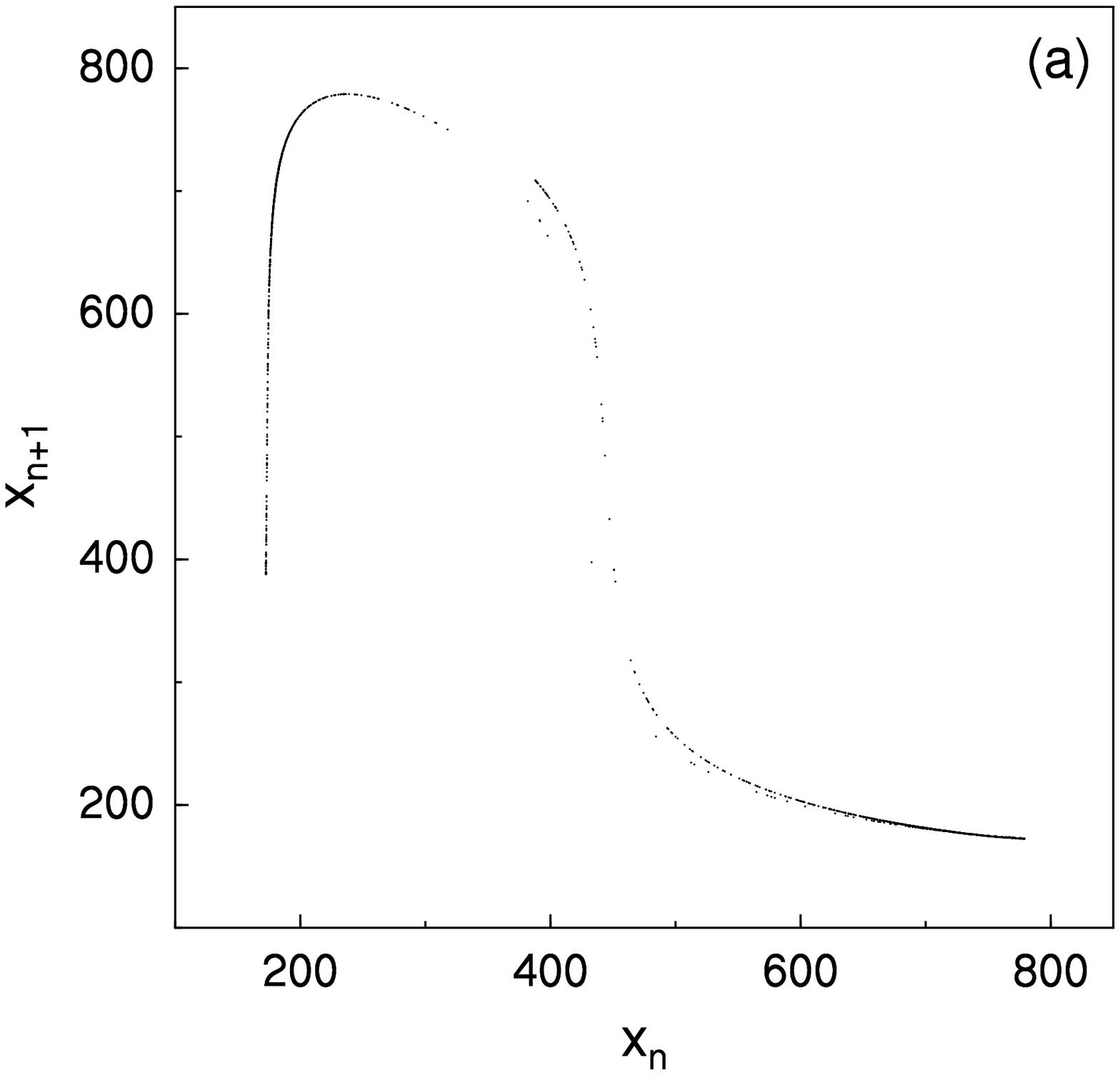}
\includegraphics[width=8.0cm]{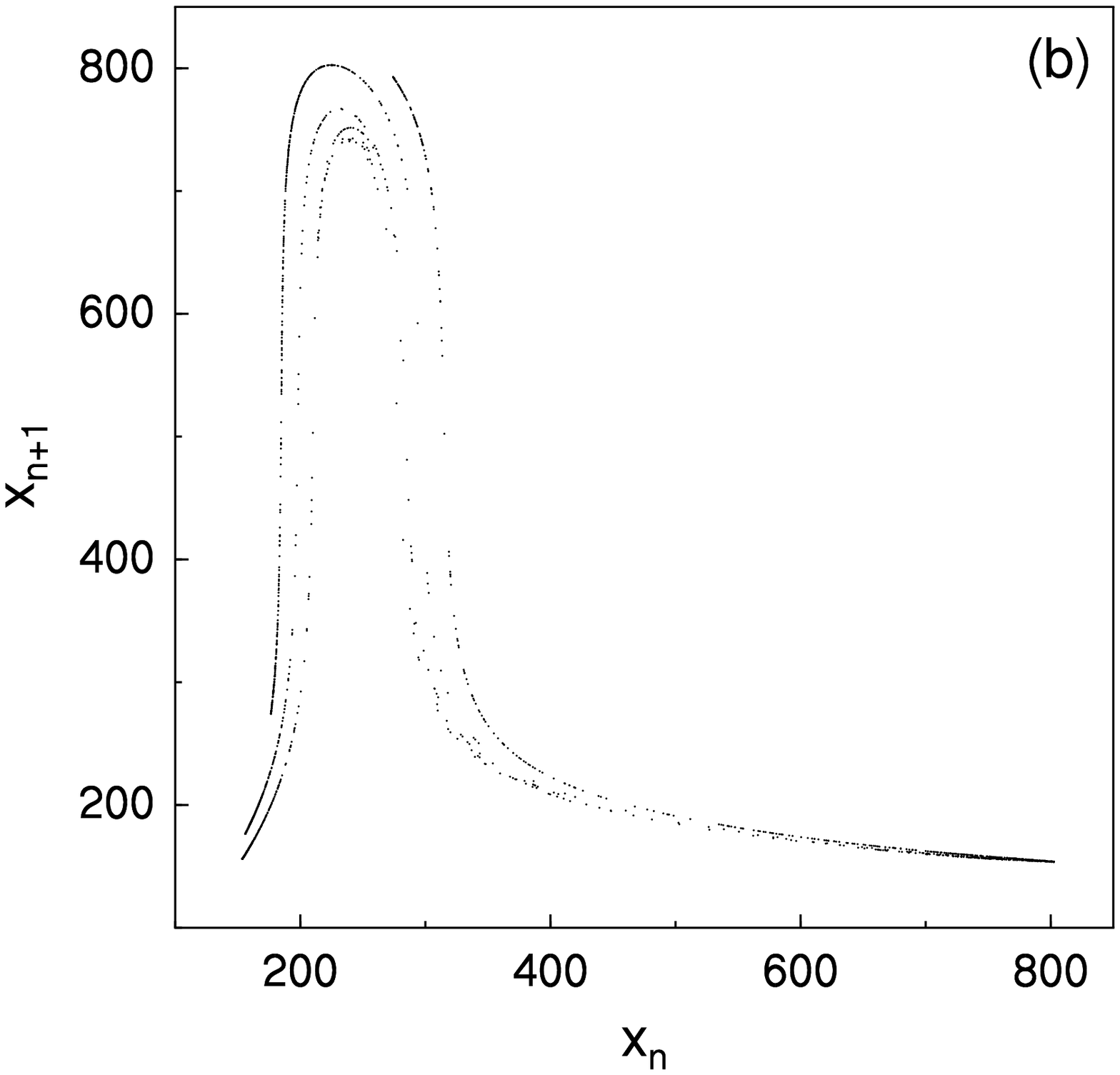} 
\caption{
Poincare maps from the  AK model for
$m=1.8$ and (a) $e=190.6$,  and (b) $e=202.7$. (Multiple folds
arise as a result of finite dissipation.)}\label{fig2}
\end{center}
\end{figure}

 As an illustration of the generic features of MMO
systems, we collect some relevant results  from our earlier study
\cite{raj00,rajesh99} on Ananthakrishna's model (AK model)
for a type of plastic instability \cite{ana82}. 
As in other MMO systems, this model also involves
disparate time scales. The bifurcation portraits ( with respect to
a primary bifurcation parameter, $e$) of the model show period doubling
sequences and their reversal, gradually changing over to period
adding sequences as the secondary bifurcation parameter, $m$,  is
increased. ( See Fig. \ref{fig1}a,b.) As  can be seen in Fig.
\ref{fig1}b, the dominant periodic orbits of $1^s$ kind constitute
the period adding sequence. (Here, we use the conventional notation of $L^s$ to  represent  period orbits in MMO systems where $L$ and $s$ correspond 
to large and small amplitude oscillations respectively.)  In the parameter space, wherein the
reversal of period doubling sequences occur,  periodic adding
sequences are finite. Periodic orbits of $1^s$ type   loose their
stability in a period doubling bifurcation as the (primary)
control parameter is increased, restabilize through a reverse
period doubling bifurcation, and are eventually annihilated in a
fold bifurcation \cite{raj00}. In this parameter range, we
have also studied the next maximal amplitude (NMA) maps
\cite{lorenz63} obtained by plotting  one maxima of the evolution
of the fast variable with the next \cite{rajesh99}. ( The NMA maps
can be regarded as a specific form of the Poincare maps.) 
These NMA maps show a near one
dimensional unimodal nature with features of sharp maximum and a
long tail. The sharpness of the NMA maps increases with the
primary bifurcation parameter for a fixed value of the secondary
bifurcation parameter. (See Fig. \ref{fig2}a and \ref{fig2}b. Also, see
\cite{rajesh99,rajesht}.) 
Several other MMO systems also exhibit
similar features.\\ 

A well studied example of MMO systems is the Belusov
Zhabotansky (BZ) reaction system. Exhaustive
theoretical/experimental studies  for the  BZ systems in two
parameter space have shown that the  NMA maps of these systems
have a unimodal structure  with a long tail and show a similar
trend as a function of the control parameters
\cite{bz,hauser96,coffman86} as in the case of AK model.
 Other MMO systems which display similar
features include lasers with a saturable absorber, autocatalytic
systems and a number of oscillating chemical reaction systems
\cite{koper95,tamosi89,hauser96,othersystems}.\\

The periodic states representative of MMO systems  have been shown
to be associated with systems exhibiting global bifurcations in the
form of approach to homoclinicity
\cite{guc90,gaspard84,glenn84,gaspard87}. Although, a variety of
tools are available for the study of bifurcations in dynamical
systems, understanding these features of MMOs has not been easy
mainly  due to the lack of adequate analytical tools to handle the
global nature of bifurcations underlying the MMOs. However, since
information regarding the  nature and properties of periodic
orbits of continuous flow systems are known to be embedded in
Poincare maps, they have been used effectively to understand the
underlying bifurcation mechanisms
\cite{lorenz63,guc90,michelin94,fang95}. For instance,
considerable insight has been obtained through the Poincare maps
derived from approach to homoclinicity
\cite{guc90,gaspard84,glenn84,gaspard87}. Since, most of the MMO
systems exhibit high dissipation, there have been attempts to
understand the dynamical behaviour of MMO system (specifically BZ
system)  by modelling its NMA maps as  one dimensional discrete
dynamical systems
\cite{coffman86,simoyi82,bagley86,ring84,pikovsky}.  \\

 The two distinct bifurcation features, namely, the 
reversal of bifurcation sequences  and the period adding sequences  
have been modelled separately using one dimensional
maps. For instance, of the many mechanisms suggested for the
reversal of period doubling sequences, a simple one involves maps
having a single critical point along with nonmonotonous dependence
of the control parameter \cite{bountis}. Another one is a unimodal
map having a negative Schwarzian derivative \cite{nusse}. Yet
another  one involves a competition between more than one critical
point in the dynamics of the map \cite{parlitz,kan92,dawson}. 
It has also been shown that the transmutation of the U-sequences to the Farey
sequences or specifically, its subset, the period adding sequences
can arise in maps with {\it more than one critical point} with
multiple parameters\cite{ringland}. However, to the best of author's
knowledge, {\it  none of these mechanisms apply} to experimental and
model MMO systems where one finds Poincare maps with unique critical point 
as well as smooth crossover from period
doubling and its reversal to Farey sequences for  a monotonous
change in a secondary bifurcation parameter. 
There has
been no explanation for this crossover as well. 
{\it Here, we show that a general class of
maps with a unique critical point satisfying some broad
constraints can show a gradual change from bifurcation diagrams
having the reversal of the period doubling sequences to period
adding sequences for a smooth change
of parameters}.\\

To achieve this objective, we first abstract common
features of MMO systems, and then impose them in the form of some
general constraints on one dimensional maps with two bifurcation 
parameters in an effort to understand the above dynamical
features of the MMO systems. ( Note that almost all MMO
systems have two or more parameters.) Using these general
constraints, we first analyse the basic mechanism responsible for
the reversal of periodic sequences of $RL^k$ type which correspond
to the dominant $1^s$ sequence in MMO systems. (Admittedly,  this
is  a much  less difficult  issue than the origin of reversal of
all the bifurcation sequences in these maps.)\\ 

We also show that
the windows of symbolic sequence $RL^k$ are the dominant windows
in the parameter space and derive scaling relations for the onset
of windows of  periodic orbits sandwiched between successive
windows of $RL^k$ sequence. The so derived {\it scaling relations
relate the slopes near the critical point and unstable fixed
point, and are unlike the usual scaling relations derived in terms
of bifurcation parameters.} Investigation of the reversal of PD
sequences in turn allows us to understand the mechanism for the
onset of the period adding sequences as well. 
{\it Our analysis
shows that the eigen value of the unstable fixed point of the map
plays a fundamental role in controlling the window widths of the
periodic orbits as a function of the control parameter in  the
family of long tailed maps.} 
Hence, the geometrical shape of the
map determined by the nature of unstable fixed point and the 
long tail structure is shown
to be the underlying factor for the period adding bifurcation
sequences. (For this reason, we refer to these general class of
maps as {\it long tailed maps}.)

\begin{figure}
\begin{center}
\includegraphics[width=8.0cm]{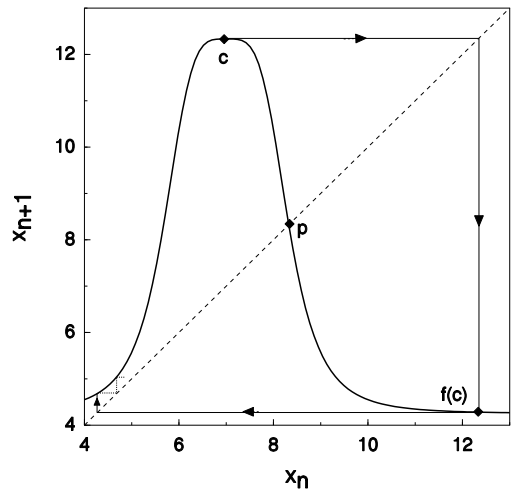}
\caption{Typical structure of the
map-L described in appendix for $\mu=4.0$ and $\xi=18.0$. The
period one fixed point(p), critical point $c$ and the first
iterate $f(c)$ are also indicated.}\label{fig3}
\end{center}
\end{figure}

We  verify these scaling relations by  constructing an example  of a two
parameter family of one dimensional maps with a geometrical shape
similar to the NMA maps of MMO systems  having a long tail. This
example map, referred to as map-L, exhibits a smooth
transition from period doubling bifurcation sequences and their
reversals to period adding sequences for a variation of a
secondary bifurcation  parameter (Appendix). We emphasise that the example map-L is primarily
introduced as an aid to verify the analytical results.
However, to pin down the ideas, it will be useful to refer to
the geometrical shape of map-L (Fig. \ref{fig3}) which is seen to be similar
to Fig. \ref{fig1}. Finally, we discuss the results and the
correspondence with the continuous flow dynamical systems
having a finite dissipation.\\

\section{Dynamics of the long tailed maps}
 In this section, , we first
attempt to extract conditions that needs to imposed on one
dimensional maps by studying MMO systems with generic features mentioned
in Introduction section  {\it without any  reference to any functional form
for the map. }   These conditions, stated below, are motivated by
the study of the long tailed maps of  MMO  systems  and are
further aided by  our detailed studies on the AK   model 
 \cite{raj00,rajesh99}. Before doing so, we first
attempt to capture the connection between periodic orbits of MMO
systems and  one dimensional maps. For the sake concreteness, we
use Shilnikov scenario where the homoclinic contact is with
respect to  a saddle focus and motivate the issues using known
facts of the AK model  studied in detail
 \cite{raj00,rajesh99,rajesht}.\\

We first identify  the equilibrium saddle fixed point of a MMO
system, as in the case of the AK model,  with
the origin of the map. In many of these experimental and model
systems, interesting dynamics arises in the region between the
first Hopf bifurcation resulting from destabilization of the fixed
point and the restabilization of the fixed point from the periodic
orbit through a reverse Hopf bifurcation. This  is identified with
back-to-back Hopf bifurcation as in the case of the AK model
also \cite{koper95,koper92,tamosi89,schi88}. 
The principal periodic orbit (PPO) born out of a Hopf
bifurcation is identified  fixed point $p$ of
the map  ( see for instance Fig. 3). Further, careful observation
of the bifurcation diagrams of MMO systems show that the complex
bifurcation sequence occurs in the region of the parameter where
the amplitude of the PPO decreases and approaches the reverse Hopf
bifurcation point as the bifurcation parameter is
increased\cite{koper95,koper92,tamosi89,raj00}.
This implies that the fixed point moves toward the origin of the
map. As the complex dynamics is terminated by  a reverse Hopf
bifurcation as the parameter is increased, the height of the map
also decreases as a function of a parameter which we identify as
the primary control parameter, $\mu$. These two statements 
can be written as two conditions 
\footnote{Even as Eq.\ref{condn1} implies that 
the whole map is coming down uniformly, Eq.\ref{condn2} 
constrains the rate of change 
to be different in different regions of the map.}:
\eqns
\frac{\d f}{\d\mu}(x) < 0, \label{condn1}\\
\vert \frac{\d f}{\d\mu}(c)\vert < \vert \frac{\partial
f}{\partial \mu}(p)\vert \label{condn2}.
\endeqns
\noindent 
where $c$ is the maximal point of the map.
We expect that these conditions are satisfied by either of the two
parameters of   the two parameter family of maps.   Without loss
of generality, we identify the parameter, $\mu$, in Eq.
\ref{condn1} and \ref{condn2} as  the primary bifurcation
parameter. (We shall use $\xi$ for the secondary bifurcation
parameter.) Here, we note that Eq. \ref{condn2}  implies that the
map becomes increasingly sharp as $\mu$ is increased. As a
consequence, the map also develops a long tail. Since, in most
cases, the variables of continuous time systems are positive, we
have taken the NMA maps to be  limited to positive values. In
Shilnikov scenario, as the orbit is reinjected close to the
origin, the orbit spiral out along the unstable manifold. Closer
the reinjection to the saddle focus, larger are the number of
small amplitude oscillations. This can be captured by ensuring an
intermittency channel in the one dimensional map. To mimic
reinjection of the phase space orbit close to the saddle focus
fixed point, we require   $f(0, \mu, \xi) =  a$ and $f(x,
\mu,\xi) \rightarrow a' < a $ for large $x$, where $a$ and
$a'$ are non zero positive constants. 
Further, since we
are interested in maps with unique critical point $c$, there are
two monotonic branches to the left and right of $c$. Further, as
the whole map is coming down as a function of $\mu$, the existence
of a intermittency channel to left of $c$, should be anticipated.
(Such a channel is actually observed in many systems
\cite{koper95,koper92,krisher93,raj00,hauser96}.) This channel is
ensured if  we impose
\eqns
f^\prime(0,\mu,\xi)  < 1 \label{condn3}\\
f^{\prime\prime}(0,\mu,\xi) > 0 \label{condn4}.
\endeqns
The above conditions (Eqs. \ref{condn1}-\ref{condn4}) are equivalent to  
the observed features (a) to (c) of MMO systems as 
mentioned in the appendix.
We shall, henceforth, refer to {\it the general class of maps
determined by these few constraints ( Eqs.
\ref{condn1}-\ref{condn4}) as long tailed maps}.\\

To fix up the preliminaries, we start with some
notations and definition.   For the sake of brevity, wherever
necessary, we will use $f(x)$ in place $f(x;\mu, \xi)$. The $n-$th
iterate of $x$ is denoted by   $f^n(x) = f(f^{n-1}(x)) = f\circ
f^{n-1}(x)$.  The iterates of $f(x;\mu,\xi)$  are given by
$\theta:\{x_0, x_1, x_2, \cdots,  x_n, \cdots \}$ with the
respective neighbourhoods given by
$\{I_0,I_1,I_2,\cdots,I_n,\cdots\} $,  where $x_n \in I_n $. A
point $q$ is $k$-periodic if $f^k(q;\mu,\xi) = q$,   while
$f^i(q;\mu,\xi) \ne q$ for $0<i<k-1$. In particular, we denote the
period one fixed point by $p$. The eigen value    of a $k$-period
cycle $q$ is $\frac{\df^k}{\dx}(q) = \lambda$.   If
$\left|\lambda\right| < 1$, then $q$ is stable. Further, we will
always choose   the initial point, $x_0$, to be    the iterate
that is closest to the critical point in the stable  periodic
window. Finally, we will deal with the nontrivial dynamics of the
map in the   interval $I = [ c_1, c_0]$, where $f(c) = c_0$ and
$f^2(c)=c_1$.  Hereafter, we refer to the parameter windows of
stable periodic orbits using the corresponding  symbolic
dynamics notation. (In this notation, to conform with the conventional
representation,  the symbol corresponding to the initial iterate
$x_0$ has been dropped. For example, we represent the entire
period   three window by
simply $RL$.)\\

A rough understanding of the reversal of the period
doubling sequences can be   obtained by examining the changes in
the stability of $p$.   For a unimodular map, Eq.(\ref{condn2})
implies that as $\mu$ is increased, as the fixed point $p$
moves towards the origin, it loses stability leading to the
first period doubling (PD) bifurcation, say   at $\mu = \mu_2$. On
the other hand, Eq.(\ref{condn1}) implies the map as a whole is
coming down which in turn means that  $p$ moves towards the
critical point, $c$, as a function of $\mu$. Once  $p$ crosses the
point of inflection, $\frac{\d f}{\d x}(p) $ increases eventually
culminating in a stable monoperiodic orbit  when $\frac{\d f}{\d
x}(p) > -1$, say at $\mu = \mu^{\prime}_2$ .   
Thus, if there are period doubling sequences beyond $\mu=\mu_2$, then 
their
reversal has to occur within the parameter range $\mu_2$ and
$\mu^{\prime}_2$.

Since the map is comes down as a function of $\mu$, there are two
possibilities for reinjection, namely $f^2(c)$  falls to  right of
a point $x = x_t$ at which $f^{\prime} (x_t,\mu,\xi) =1$ and
$f^2(c)$ falls to its left. The latter clearly corresponds to  the
realization of  period adding  sequences. 
Any further
increase in the parameter creates two more fixed points on the
ascending branch of the map. 

\begin{figure}
\begin{center}
\includegraphics[width=8.0cm]{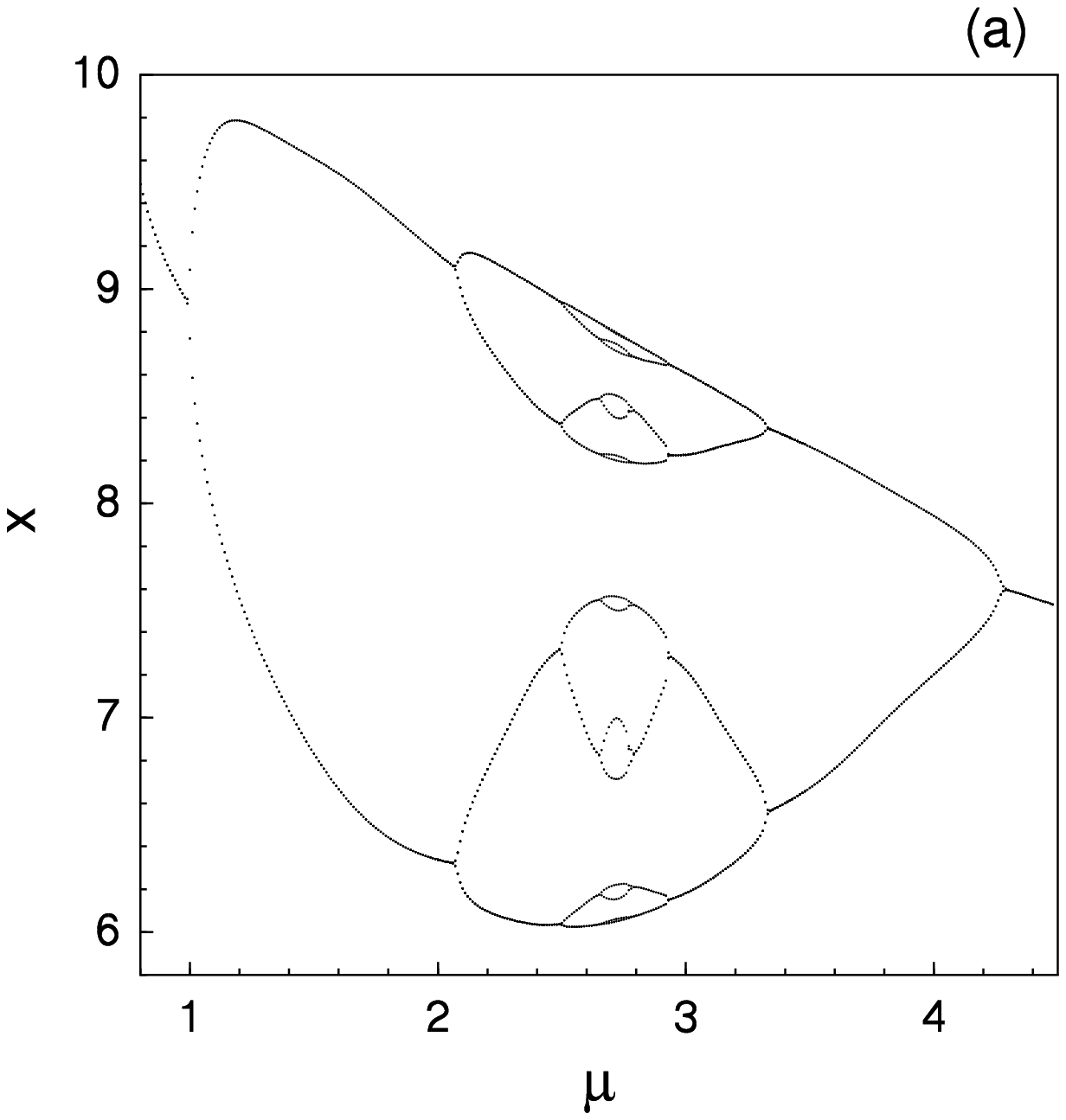}
\includegraphics[width=8.0cm]{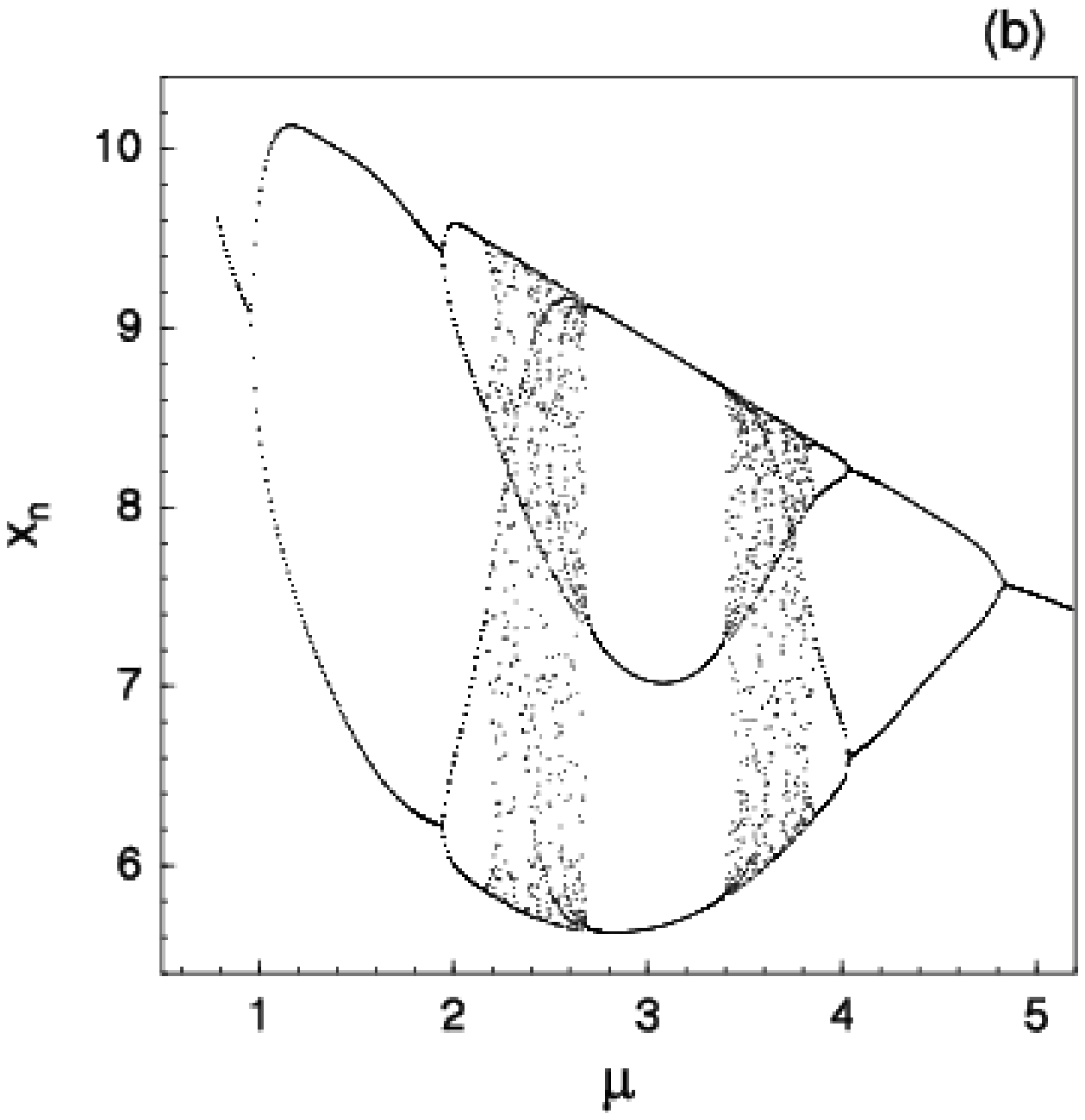} 
\caption{
Bifurcation diagrams for the Map-L. (a) Cascading period
doubling with a bubble structure for  $\xi = 11.415$, and (b) Bubble structure with 
dominant period three window for $\xi=11.43$. }\label{fig4}
\end{center}
\end{figure}

\subsection{Reversal of bifurcation sequences}

Before we proceed further, we  give below some  derivatives which   
will be
used  to establish subsequent  results. Using the  chain rule for 
iterates
of maps, we have

\eqns \frac{\d f^n}{\dx}{(x_0)} &=& \prod_{i=0}^{n-1}
\frac{\df}{\dx} (f^i (x_0)). \label{lambdan}
\endeqns

\noindent Using  Eq. (\ref{lambdan}), we can write
\eqns \nonumber
\frac{\d^2 f^n}{\dx^2}{(x_0)}&=&\frac{\d}{\dx} \left( \frac{\d
f^n}{\dx}(x_0)\right),\\
 &=&\sum_{k=0}^{n-1} \left( \frac{\frac{\d^2 f}{\dx^2} 
(f^{k}(x_0))}{\frac{\df}{\d x}(f^k(x_0))}\right) \prod_{j=0}^{n-1} 
\frac{\df}{\dx} (f^j
 (x_0)) \label{curvature}.
\endeqns
\noindent
Similarly, the parameter dependence of the iterates lead to
\eqns
   \frac{d f^n}{d\mu}(x) &=& \frac{\d f}{\d\mu}(f^{n-1}(x)) +  \frac 
{\d f}{\dx}(f^{n-1}(x)) \frac{\d f^{n-1}}{\d
   \mu}(x).\label{variation}
\endeqns

 In order to understand the mechanism of reversal of
period doubling sequences    exhibited by the unimodal long tailed
maps, we restrict our discussion to  a  simpler problem namely,
the reversal of   period doubling of the  dominant  sequences of
the type $RL^k$.   (We will soon  argue that these type of
periodic orbits are the dominant ones.)

\begin{figure}
\begin{center}
\includegraphics[width=8.0cm]{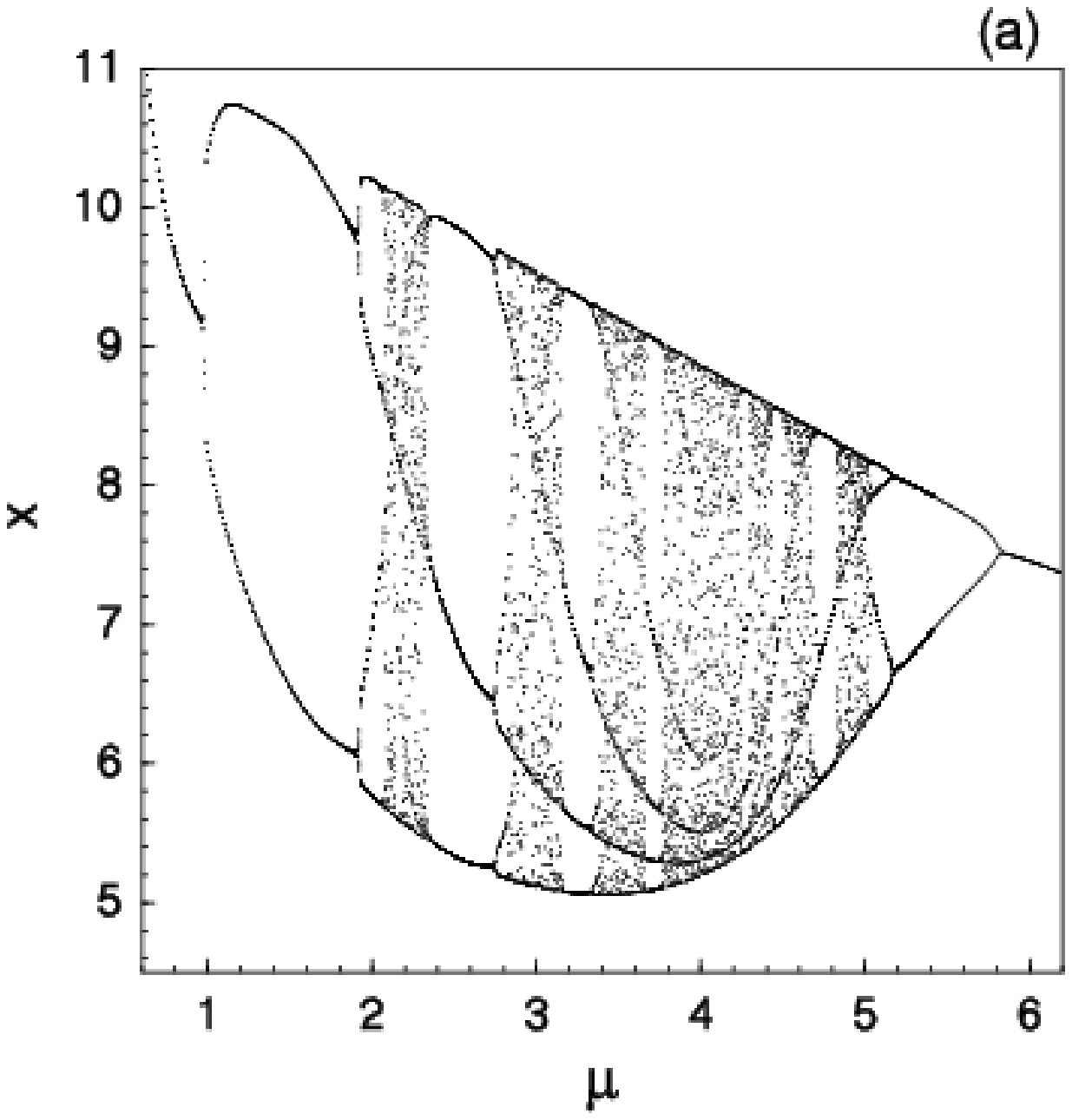}
\includegraphics[width=8.0cm]{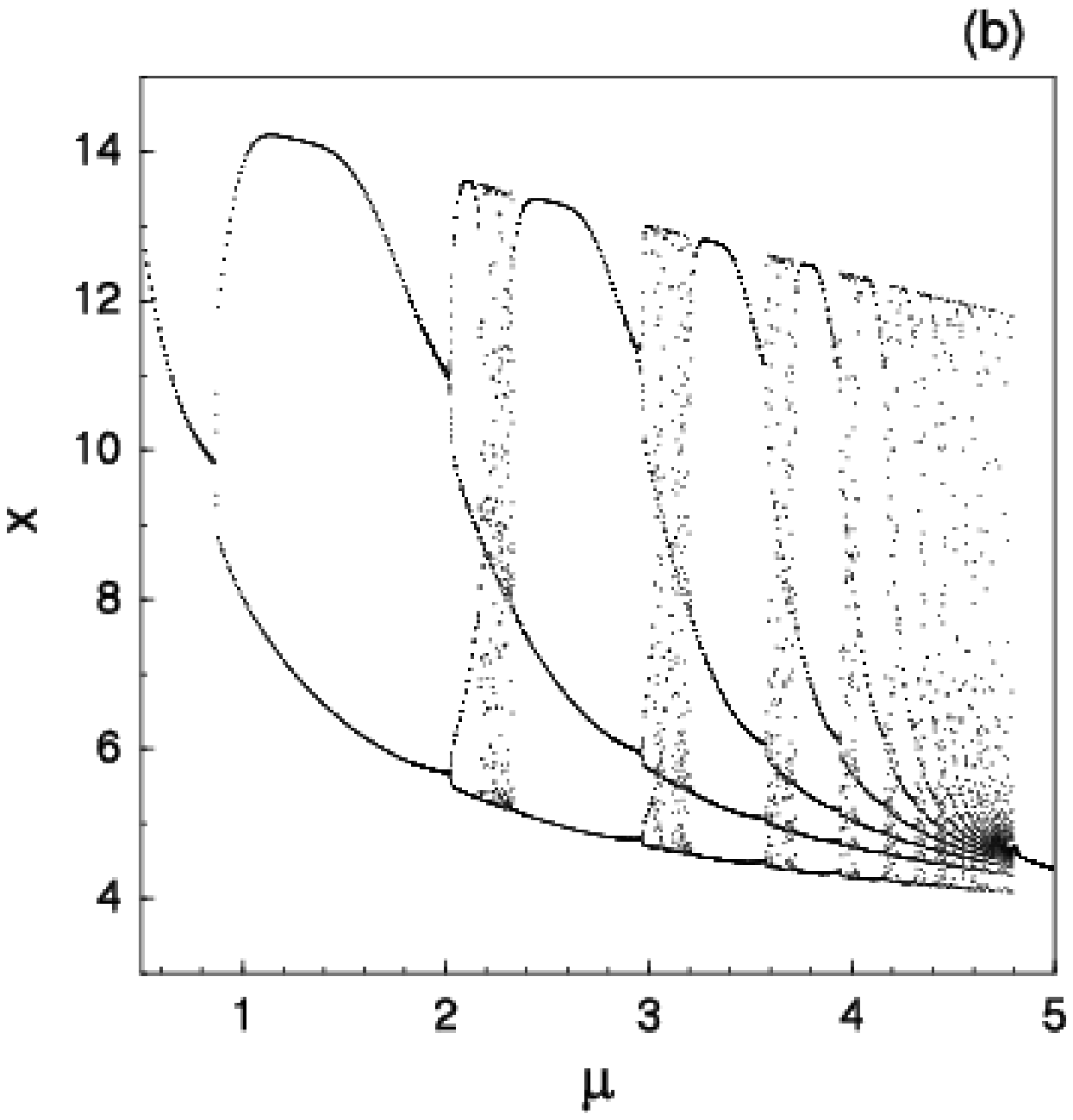} 
\caption{ The bifurcation diagrams for Map-L. (a) period adding
sequence coexisting with reversal of period doubling sequences  at
$\xi=12.8$, and (b) period adding sequence with fold bifurcation
at the accumulation point of $RL^n$ periodic windows for
$\xi=18.0$. Note the similarity with Fig. \ref{fig1} b.}\label{fig5}
\end{center}
\end{figure}

To pin down the ideas, consider the map-L which satisfies the
requirement of the long tailed maps. The dominance of the
$RL^k$ periodic orbits are well visualised in the bifurcation
diagrams of map-L (Fig. \ref{fig5}a). A prominent feature of the
bifurcation diagram is the presence of relatively dark bands of
iterates ( corresponding to superstable orbits) that run across
the bifurcation diagram connecting successive $RL^k$ periodic
windows. (Similar features are routinely observed in bifurcation
diagrams of MMO systems also.) These bands arise as a consequence
of the increased stability of the iterates closer to the critical
point of the map. It is clear that these bands share a similar
sequential structure of iterates as the $RL^k$ sequence,
specifically the superstable point of the periodic orbit lies on
the envelope of these bands.
In the following analysis and later, we utilise this contiguous 
nature of envelope of bands to determine the parametric 
dependence of iterates. \\

Consider the $n$ iterates of $k + 2$ periodic cycle which consists
of one visit to the right of the critical point, $c$, and $k$
visits  to the monotonically increasing arm   of the map. We
examine the reversal of the sign   of Eq.(\ref{variation}) as
$\mu$ is increased for periodic orbits of the form $RL^k$ as the
reversal of the periodic sequence is assured if $\frac{\d
f^2}{\d\mu}{(x_0)}$ changes sign at $\mu = \mu_{r_2}$. To do this,
first consider the second iterate ( $n =2$) of
Eq.(\ref{variation}). Using $x_0 = c$, we get
\eqns \frac{\partial
f^{2}}{\partial \mu}(c)=\frac{\partial f}{\partial \mu}(f(c))+
\frac{\partial f}{
\partial x}(f(c))\frac{\partial f}{\partial
\mu}(c).\label{variation2}
\endeqns
Equation (\ref{condn1}) implies that the first term on the RHS of
Eq.(\ref{variation2}) is negative while the second term is
positive since $\frac{\partial f}{  \partial x}(f(c))$ and
$\frac{\partial f}{\partial \mu}(f(c))$  are negative. However, initially,
$\vert \frac{\partial f}{\partial x}(f(c)) \vert$ {\it is small as
$f(c)$ refers   to the tail part}, and hence there is a region of
$\mu$ for which $\vert \frac{\partial f}{\partial \mu}(f(c))\vert
> \vert \frac{\partial f}{\partial x}(f(c))\frac{\partial
f}{\partial \mu}(c)\vert$ and thus, LHS of Eq.(\ref{variation2})
is negative. As $\mu$ is increased 
( as $f(c)$ moves up
towards $c$),  
$\vert \frac{\partial f}{\partial x}(f(c)) \vert$ increases and $\frac{\partial f^{2}}{\partial \mu}(c)$ turns
positive, say at $\mu = \mu_{r_2}$. (Note that the value of
$\mu_{r_2}$  is in principle a function of  $\xi$. For example, in the
case of map-L, for $\xi = 11.0$, $\frac{\partial f^{2}}{\partial
\mu}(c)$ changes sign at $\mu \sim 2.2$.) Beyond
$\mu=\mu_{r_2}$,$\frac{\partial f^{2}}{\partial
\mu}(c)$  increases with $\mu$.\\

Now, consider evaluating Eq.(\ref{variation}) for
successive $n> 2$. Consider,
\begin{equation}
\frac{\partial f^{3}}{\partial \mu}(c) = \frac{\partial
f}{\partial \mu}(f^{2}(c)) + \frac{\partial f}{\partial
x}(f^{2}(c))\frac{\partial f^2}{\partial \mu}(c)
\label{variation3}
\end{equation}
\noindent For periodic orbits of type $RL^k$,  $\frac{\partial
f}{\partial x}(f^{2}(c)) > 0$, and  from Eq. (\ref{condn1}),
$\frac{\partial f}{\partial \mu}(f^2(c)) < 0$. Further,  from
the previous discussion,  since $\frac{\partial
f^{2}}{\partial \mu}(c)$ is zero at $\mu = \mu_{r_2}$, LHS of
Eq.(\ref{variation3}) negative at this value. As $\mu$
is increased, $\frac{\partial f^{2}}{\partial \mu}(c)$ turns positive
and the second term on the RHS of Eq.(\ref{variation3}) is an
increasing function of $\mu$. Thus, the LHS which is negative for
$\mu < \mu_{r_2}$ becomes positive at some value $\mu = \mu_{r_3}
> \mu_{r_2}$. In addition, note that $\vert \frac{\partial
f^{3}}{\partial \mu}(c)\vert > \vert\frac{\partial f^{2}}{\partial
\mu}(c)\vert$ around $\mu = \mu_{r_3}$. Similar arguments can be
used to show that $\frac{\partial f^{n}}{\partial \mu}(c)$, which
is negative for $\mu < \mu_{r_{n-1}}$ will change sign  at $\mu =
\mu_{r_n}$. Thus, the reversal of successive periodic orbits occur
at increasing values of parameter given by \eqns \nonumber
    \mu_{r2} < \mu_{r3} < \mu_{r4} < ...
\label{order}
\endeqns
\noindent (Note that these results are largely a consequence of
Eqs.(1) and (2).) These results can be verified from the
bifurcation portraits of map-L shown in Fig. 5a. Changes in sign
of $\frac{\partial f^{n}}{\partial \mu}(c)$, manifest as changes
in the sign of the slope of the envelope of superstable iterates
with parameter.  It is clear that lower outermost envelope
corresponding to the third iterate of critical point changes the
sign of   slope first, signalling the onset of
reversal of the periodic sequence.\\

A natural consequence of the reversal of periodic orbits
is the presence of isolated bifurcation curves or the isolas. In
the following, we trace the origin of  of isolas using  the
first two constraints (Eqs. \ref{condn1},\ref{condn2}) on the
general class of maps. Again, we restrict our attention to
periodic orbits of the form $RL^k$. Consider using $ x_0 = c$ in
Eq.(\ref{curvature}) for an allowed $k$-periodic $RL^{k-2}$, we
get 
\eqns 
\frac{\d^2 f^k}{\dx^2}(c) &=& \frac{\d^2 f}{\dx^2}(c) \nonumber
\prod_{n = 1}^{k-1} \frac{\d f}{\dx}(x_n)\\ \nonumber
&=&\frac{\d^2 f}{\dx^2}(c) . \frac{\d f}{\dx}(f(c)) .
\prod_{n=2}^{k-1} \frac{\d f}{\dx}(x_n) \label{curvature1}
\endeqns
 We first note that excepting $\frac{\d f}{\dx}(f(c))$
and $\frac{\d^2 f}{\dx^2}(c)$, all other terms are positive. Thus,
as the visits are to the left of $c$, the sign of $
\frac{\d^2 f^k}{\dx^2}(c)$ is positive for all allowed values of
$\mu$. This positive curvature of the map and its
parametric dependence, $ \frac{\d f^k}{\d \mu}(c) < 0$ 
for $\mu < \mu_{r_k}$,  implies that a $k$-periodic orbit is
created due to a tangent bifurcation as $\mu$ is increased.
For further increase in $\mu$, $f^k(c)$ comes down, triggering  a
period doubling bifurcation making the periodic orbit unstable.
Increasing  $\mu$ beyond $\mu_{r_k}$ reverses the period doubling
since $\frac{\d f^k}{\d \mu}(c)$ becomes positive, ie., the region
around $c$ of $f^k(x)$ moves up eventually restabilizing the
$k-$th periodic orbit. Further increase in $\mu$ destroys  the
stable periodic orbit in another tangent bifurcation (see Fig. 5b).
The creation of a periodic orbit, its destabilization   and
restabilization, followed by its destruction constitutes an
isolated bifurcation curve (isola). This happens for each  of the
allowed periodic orbit of the form $RL^k$. 
From these arguments, it is clear that   the isola corresponding
to the smallest allowed period orbit contains all   isolas of the
allowed periodic orbits of larger $k$, with the smallest  one
being the maximum periodic orbit. In addition, there exists a
maximum periodic orbit  of $RL^k$ that is allowed.  In the case of
map-L, it is clear that the value at which different iterates show
reversal in sign $\frac{\partial f^{n}}{\partial \mu}(c)$ are
different as can be seen from Fig. 5a and the reversal of all the
allowed periodic orbits of $RL^k$ type occurs  beyond  $\mu \sim
4.05$ ($\xi=8.0$) and the maximum allowed periodic  sequence
period  five ( $k =3$). Indeed, similar isola features  are well
documented in bifurcation diagrams of MMO
systems\cite{koper95, raj00}.\\

It is possible to extend these arguments to some other
sequences of periodic  orbits as well.  For example, consider
periodic orbits of $RL^{k}R^{2m}$ type. Similar arguments as
before show that $ \frac{\d^2 f^{k+2(m+1)}}{\dx^2}(c)$ is positive
and $ \frac{\partial f^{k+2(m+1)}}{\partial \mu}(c)$ changes sign
from negative to positive, hence forming isola structures.
However, there are other kinds of periodic  orbits in the allowed
symbolic sequence (for example, $RL^kR^{2m+1}$),  whose reversal
is interrupted by the homoclinic points. Extending the above type
of arguments to  such cases will  depend on the nature
of the sequences and appears difficult.\\

As discussed in the introduction, one  important feature
of MMO systems is the dominance of the parameter windows of stable
periodic   orbits of $RL^k$ type. Using Eq.(\ref{curvature1}),  we
show that the width  of   the parameter windows  of these stable
periodic orbits depend critically on the structure of the map as
determined by Eqs.(\ref{condn1}) and (\ref{condn2}).   In
particular, we will show that the ratio of the   widths of
successive periodic windows of the $RL^k$ type is decreasing.
Consider Eq.(\ref{curvature1}) for $ k $ and $k +1$ stable
periodic orbits. In comparison with the $k$-th periodic orbit,
there is an extra term in the product arising from an iterate
falling on the monotonically increasing arm. Since the $\mu$ value
of the $(k + 1)$-th periodic orbit  is larger than that for $k$-th
one, which in turn implies $f(c)$ has moved towards $c$,  we have
\eqns 
\prod_{n=2}^{k-1}\frac{\d f}{\dx}(f^{n}(c)) < \prod_{n=2}^{k}\frac{\d f}{\dx}(f^{n}(c)) .
\endeqns
Further,  since $f(c)$ moves towards the critical point,
$\vert\frac{\d f}{\dx}(f(c))\vert$ increases. Noting that
$\frac{\d^2 f^k}{\d x^2}(c)$ is negative, whose magnitude
increases with $k$ ( or $\mu$), it implies that  $f^{k + 1}(x)$
exhibits a sharper minimum around $c$ than  that for the  $k$-th
periodic orbit. Assuming that the extent of the neighbourhood
around $f^k(c)$ can be approximated by $F^k \sim \frac{\d^2
f^k}{\d x^2}(c) (x - c)^2/2$,  the window width of the $k$-th
periodic orbit can be approximated by the product of $F^k$ and
$\frac{\d f^k}{\d \mu}(c)$. Since $\vert F^k \vert < \vert
F^{k + 1} \vert$ and $\left |\frac{\d f^k}{\d \mu}(c) \right | <
\left |\frac{\d f^{k + 1}}{\d \mu}(c) \right |$, it clear that the
window width of the $k$-th periodic orbit is larger  than that of
$(k + 1)$ periodic orbit.
Since the value at which the reversal of PD sequence of
the type $RL^k$ occurs depend on the secondary bifurcation
parameter $\xi$, it is possible that there may not be a
change in the sign of $\frac{\d f^k}{\d \mu}(c)$ beyond a certain
value of $k$. ( Note that if $\frac {\partial f^k}{\partial \mu}$
does not change sign, then, there cannot be a change in sign for
$m > k$. ) Thus, only a few periodic sates of $RL^k$ would be
seen. However, for the  period adding sequences, the number of
states of $RL^k$ can go  unbounded.
 For this case, a channel at small $x$ should open up.
The onset of a channel can be anticipated based on   Eq.
(\ref{condn1}) which implies that the graph of the map comes down
with $\mu$, and $f(0,\xi,\mu)=a$, i.e., nonzero value as the
origin of the map.  A power series representation for the
monotonically increasing part with conditions $f^\prime(0,\mu,\xi)
< 1$ and $f^{\prime\prime}(0,\mu,\xi) > 0$ ensures that there is
one point to left of $c$ that has $f^\prime(x,\mu,\xi) = 1$ for
all values of the parameter. Then, for $RL^k$ periodic sequences,
the window
widths of periodic orbits of the type $RL^k$  decreases slowly. \\

\subsection{Scaling relations for periodic orbits}

Having argued for the predominance of the periodic
orbits of sequence of $RL^k$ type in the parameter space,   we
show that periodic windows contained in the chaotic region
bounded by the windows of  periodic orbits denoted by      $RL^k$
and $RL^{k+1}$ have    much smaller parameter width   compared to
that of $RL^k$ and $RL^{k + 1}$.    Numerical results on the map -
L are also   presented in support of the analytical result.\\

The PA sequence also manifest in NS-unimodal maps
\cite{geisel81} as a part of MSS sequence\cite{mss}. In the MSS
sequence, between any successive  $RL^{k-1}$ and $RL^k$ windows,
there exist windows of  allowed sequences of the type $RL^k[S]$,
where $[S]$ is a     sequence consistent  with the allowed symbol
sequences\cite{post91}, i.e.,
\begin{eqnarray}
RL^{k-1} \prec \cdots \prec RL^k[S] \cdots \prec RL^k. \nonumber
\end{eqnarray}
Considering only the increasing  behaviour of  long tailed  maps,
we assume that the MSS   ordering corresponding to the periodic
orbits of  NS-unimodal maps   may be applied  for   these types of
maps also.

Let the bifurcation values $\tilde{\mu} _k$ correspond
to the last generation $k$-period orbit in the  MSS sequence,
i.e.,  for $\mu >  \tilde{\mu}_k$ no more cycles of period $k$
are present. Thus,   the parameter values are ordered in the
following way:
\begin{eqnarray}
\nonumber           \mu_3 = \tilde{\mu}_3 <  \tilde{\mu}^h_3 <
\tilde{\mu}_4 <  \tilde{\mu}^h_4 < \tilde{\mu}_5 < \cdots
\nonumber
\end{eqnarray}
where $\tilde{\mu}_k^h$ refers  to  degenerate homoclinic bifurcation
\cite{gardini} (wherein $f^k(c)=p$) of the periodic  point $p$
between $k$-th
and $(k+1)$-th periodic orbits.\\

\begin{figure}
\begin{center}
\includegraphics[width=8.0cm]{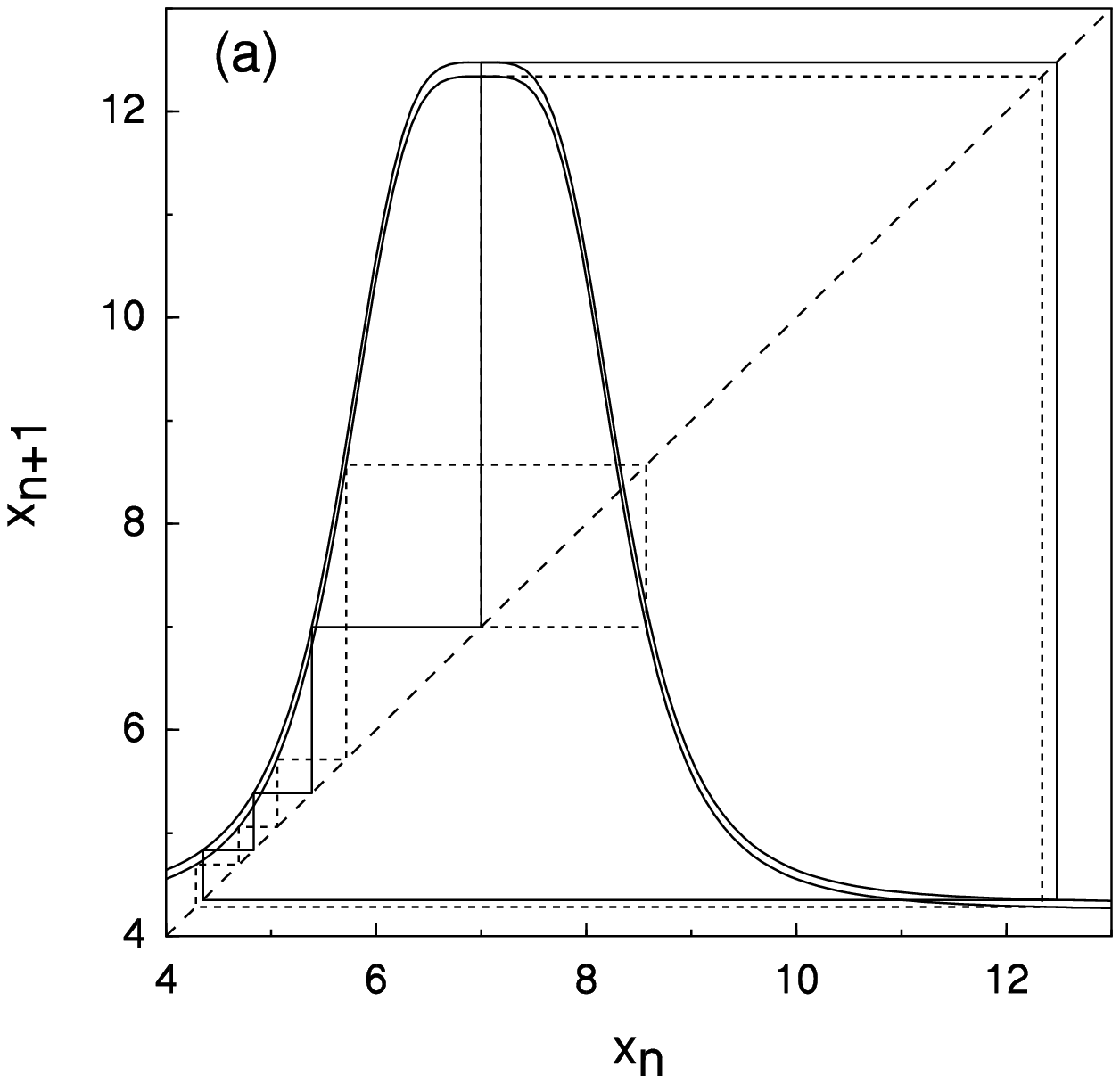}
\includegraphics[width=8.0cm]{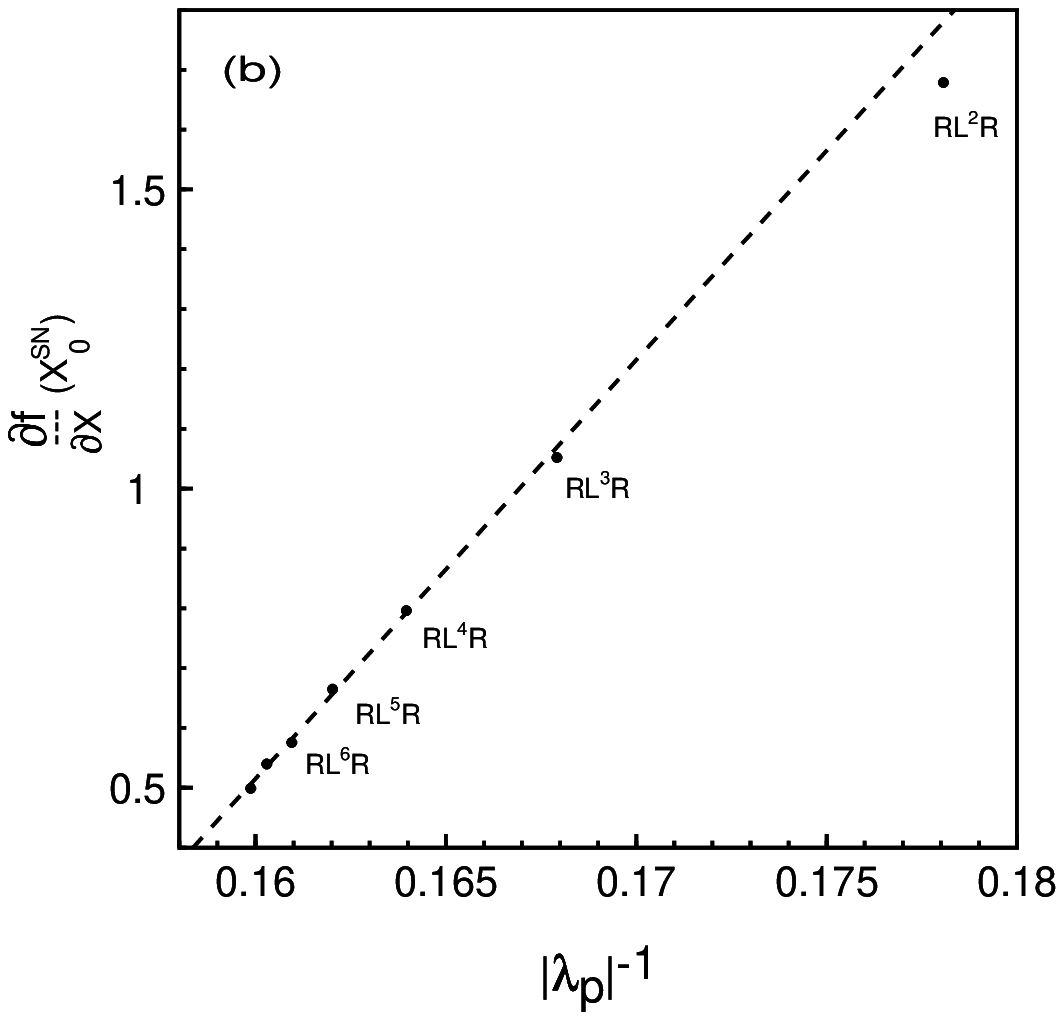} 
\caption{
(a) Map-L and the periodic orbit of form $RL^3$ and $RL^4R$ for
$\mu=3.7831$ and $\mu=3.99$ respectively ($\xi=18.0$). (b) Verification of 
the scaling of  $\frac{\partial
f}{\partial x}(x^{SN}_0)$ with $\vert\lambda_p\vert^{-1}$ for 
$RL^nR$ periodic sequences obtained from the example Map-L. The dashed line is drawn
for a linear fit.}\label{fig6}
\end{center}
\end{figure}

Using a degenerate  homoclinic    bifurcation  as the
boundary, the parameter space for all the periodic windows between
$RL^{k-1}$ and $RL^k$ ( of period $k + 1$ and $k +2 $) can be
divided into two regions, as  $R_I : [\tilde{\mu}_{k+1},
\tilde{\mu}_{k+1}^h]$ and $R_{II} : [\tilde{\mu}_{k+1}^h,
\tilde{\mu}_{k+2}]$. Consider the periodic window corresponding to
the simplest symbolic sequence in the regions $R_I$  where stable
periodic windows are of the type $RL^kR[S]$. Within this sequence,
the most dominant ( in the sense of the MSS sequence) periodic
sequence  is $RL^kR$ which has one period higher than the basic
sequence $RL^k$. Using Eq.(\ref{lambdan}), for   a $k-$th periodic
orbit ( $f^k(x_0) = x_0$), the eigen value of the periodic orbit,
$\lambda$, is given by,
\begin{equation}
\label{stability}
                \lambda =    \prod_{n=0}^{n=k-1} \frac{\d f}{\d x}(x_n)
\end{equation}
\noindent and for stability $ \left | \lambda \right | \le 1.0$.
(For the period one fixed point, we have
 $ \lambda_p = \left |\frac{\d f}{\d x}(p)\right | < 1.0 $  for
$\mu \le \mu_2$.) Within the first order, we approximate all
iterates falling in the  neighbourhood $U$ of $p$ as having the
same slope as  at $p$. In the case of long tailed maps, 
owing to the sharp nature of longtailed maps,  $\left |
\frac{d f}{d x}(p) \right |$ is large, for $\mu
> \mu_3$.  For the sake of clarity, consider
the sequence $RL^kR$ which is part of the sequence $RL^kR[S]$ in
$R_I$. Clearly, the structure of the map ensures that one iterate
of this sequence belongs to $U$ (see for example map-L in Fig. 6a
). In the case of $RL^kR$ sequence, $f^{k+2}(x_0)$ falls in
neighbourhood $U$ of $p$. Using Eq. (\ref{condn2}) for stability of
the periodic orbit,
\begin{eqnarray}
\left| \frac{\d f}{\d x}(x_0) \frac{\d f}{\d x}(x_{k+2})
\prod_{n=1}^{k+1} \frac{d f}{d x}(x_n)\right| \leq 1.0 
\end{eqnarray}
\noindent where $x_0$ is an initial point  chosen to be  in the
neighbourhood of the critical point for reasons that will  become
clear soon. Since long tailed maps are  sharply peaked (
Eqs.(\ref{condn1}) and (\ref{condn2}) ), almost all the iterates
are trapped in the intermittency   channel like region near the
origin of the map. (This point is best illustrated by considering
the concrete example of  map-L shown in Fig. 6a, where $RL^3$ and
$RL^4R$ are shown.) Thus, for large $k$, noting that iterates other 
than
$x_0$   and $x_{k+2}$ change   only marginally, for stability of
the periodic orbit, we have
\begin{eqnarray}
C \left| \frac{\d f}{\dx}(x_0) \frac{\d f}{\dx}(x_{k+2}) \right| \leq 
1.0 \nonumber
\end{eqnarray}
\noindent
where $C$ ( $= \prod_{n=1}^{k+1} f^\prime(x_n)$) is a constant factor.
Since  $x_{k+2}$ falls in the neighbourhood of $p \in U$,
$ \left | \frac{\d f}{\dx}(x_{k+2}) \right | \sim \vert\lambda_p\vert ( 
\gg 1.0) $.
Hence,
\begin{eqnarray}
\left| \frac{\d f}{\dx}(x_0) \right| \leq  \left[ 
C\vert\lambda_p\vert\right]^{-1}.
\end{eqnarray}
Recall that for any periodic orbit, one iterate is always located
in the neighbourhood of the critical point in the interval $[
x^{PD}_0, x^{SN}_0]$, where $x^{SN}_0$ and $x^{PD}_0$ correspond
to the onset of the periodic orbit through a   saddle node
bifurcation at $\mu^{SN}_n $ and the destabilization of the
periodic  orbit   through a period doubling bifurcation at
$\mu^{PD}_n$. As the parameter   is increased from  $ \mu =
\mu^{SN}_n$, the iterate $x_0$ can be regarded   as traversing
from $x^{SN}_0$ to $x^{PD}_0$. Then, Eq. (14) implies  that   for
the stability of the periodic orbit $RL^kR$,   the iterate ($x_0$)
is restricted to  a smaller extent of the  neighbourhood of the
critical point than that corresponding to $RL^k$ in order to
compensate for the increase of $\left | \frac{\d f}{\dx}(x_{k+2})
\right| \sim \vert\lambda_p\vert $. Assuming a uniform change in $x_0$ 
with respect to the
parameter $\mu$ as it traverses from $x^{SN}_0$  to  $x^{PD}_0$
and   using the fact that $\lambda_p \gg 1.0$,  the parameter
windows of   $RL^kR$ is smaller by a factor $\lambda_p$ than that
for $RL^k$.   Similar arguments can be used to show that any
sequence with $2m + 1$ extra   visits to the right than  $RL^k$
would have a window width smaller by a factor   $\lambda_p^{-(2m +
1)}$. (The assumption here is that these visits to the right   are
close to $p$.) Since any $RL^kR[S]$ sequence consists of $2m + 1$
right visits over and above the basic sequence $RL^k$, the width
of   this periodic window is vanishingly small for large $m$.
Thus,
\begin{eqnarray}
\left| \frac{\d f}{\dx}(x_0) \right| \leq  C^{-1}
\left|\lambda_p^{-(2m+1)}\right|. \label{scaling}
\end{eqnarray}

More specifically,  Eqs. (13) and (14) allows us to get a
scaling relation between the onset of   the periodic windows  of
type $RL^kR^{2m+1}$ which can be verified by using   the map-L.
For this we fix the control   parameter, $\mu$ at the fold
bifurcation points of the periodic   windows so that the equality
holds in  Eq. (14). Under this condition, Eq.(14) (
for $m=0$) implies that $\left| \frac{\d f}{\dx}(x^{SN}_0) \right|
\lambda_p = C^{-1}$ which is almost  independent of $k$ in $RL^kR$. To 
verify
this, we have plotted $\left| \frac{\d f}{\dx}(x^{SN}_0) \right| $
as a function of $ \vert\lambda_p\vert^{-1}$  which exhibits
linear scaling behaviour for  the onset values of $RL^nR$ windows
using the map-L (Fig. 6b). In a similar fashion, we have shown a
plot of $\frac{\d f}{\d x}(x^{SN}_0)$  as a function of
$\vert\lambda_p\vert^{-(2m+1)}$  for various $m$ values
corresponding to   periodic orbits of $RL^2R^{2m + 1}$  type (Fig.
7). Note that, even though  large $k$ approximation is used, the
scaling relations work well when $k$ is not large (Figs. 6b and 7).\\

\begin{figure}
\begin{center}
\includegraphics[width=8.0cm]{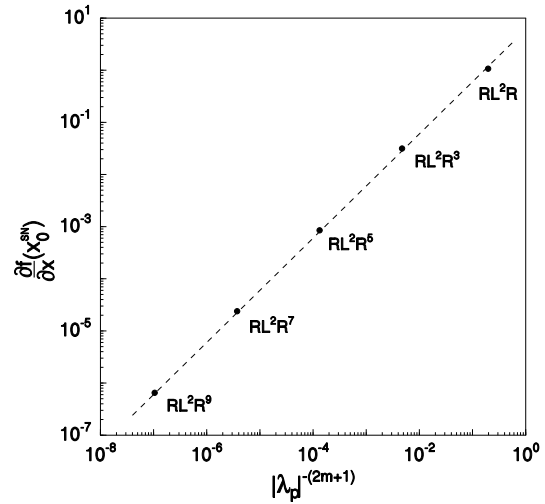}
\caption{ Verification of the
 scaling of  
 $\frac{\partial f}{\partial x}(x^{SN}_0)$ with 
 $\vert\lambda_p\vert^{-(2m+1)}$  
 for $RL^2R^{2m+1}$ periodic sequences obtained from example Map-L.}
\end{center}
\end{figure}\label{fig7}

In the region $R_{II}$,   the allowed periodic orbits
are made up of  the sequence  $RL^kR^2[S]$ which has at least two  
more iterates to the right than the corresponding $RL^k$ orbit. (See Fig. 8a.)
The smallest allowed periodic orbit  in this region is $RL^kR^2$
which has   exactly two  iterates more than the periodic orbit
$RL^k$ both of which lie in the   neighbourhood $U$ of $p$. ( Note
that even though the actual distance   of these iterates may be
substantially away from $p$, due to sharpness   of the map, the
slopes at these points can be approximated by the slope at $p$.)
Thus, approximating the derivatives of the map for these  two
iterates falling in the  neighbourhood $U$ of $p$ to be
$\lambda_p$,    the  eigen value of  the periodic orbit $
RL^kR^2$ is seen to increase by a factor $\lambda_p^2$ compared to
$RL^k$ orbit. Hence, we have
\begin{equation}
            \left| \frac{\d f}{\d x}(x_0) \right| \leq 
C^{-1}.\left|\lambda_p\right|^{-2}.
\end{equation}

Following arguments presented for $R_I$, we see that
the iterate $x_0$ is restricted to a smaller extent around the
critical point by a factor $\lambda_p^{-2}$ which in turn leads to
the width of $RL^kR^2$ being smaller by the same factor compared
to $RL^k$. Using the equality, the above relation has been
verified numerically  for the map-L with the control parameter,
$\mu$, kept at the   fold bifurcation point.  Figure 8b shows the
scaling for the onset of the periodic orbit   $RL^kR ^2$ with 
$\lambda_p^{-2}$ which confirms the analytical result.\\

\begin{figure}
\begin{center}
\includegraphics[width=8.0cm]{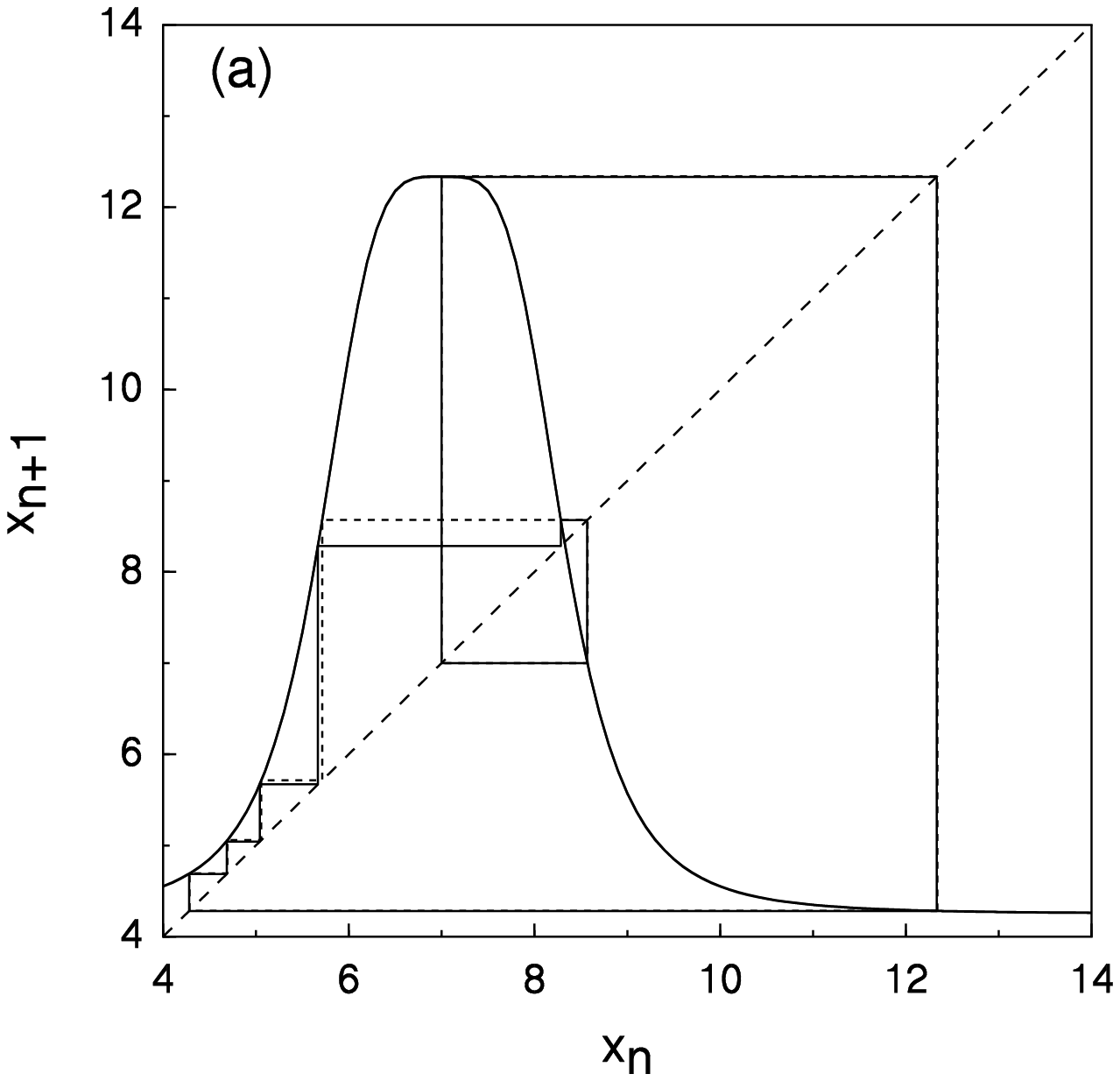}
\includegraphics[width=8.0cm]{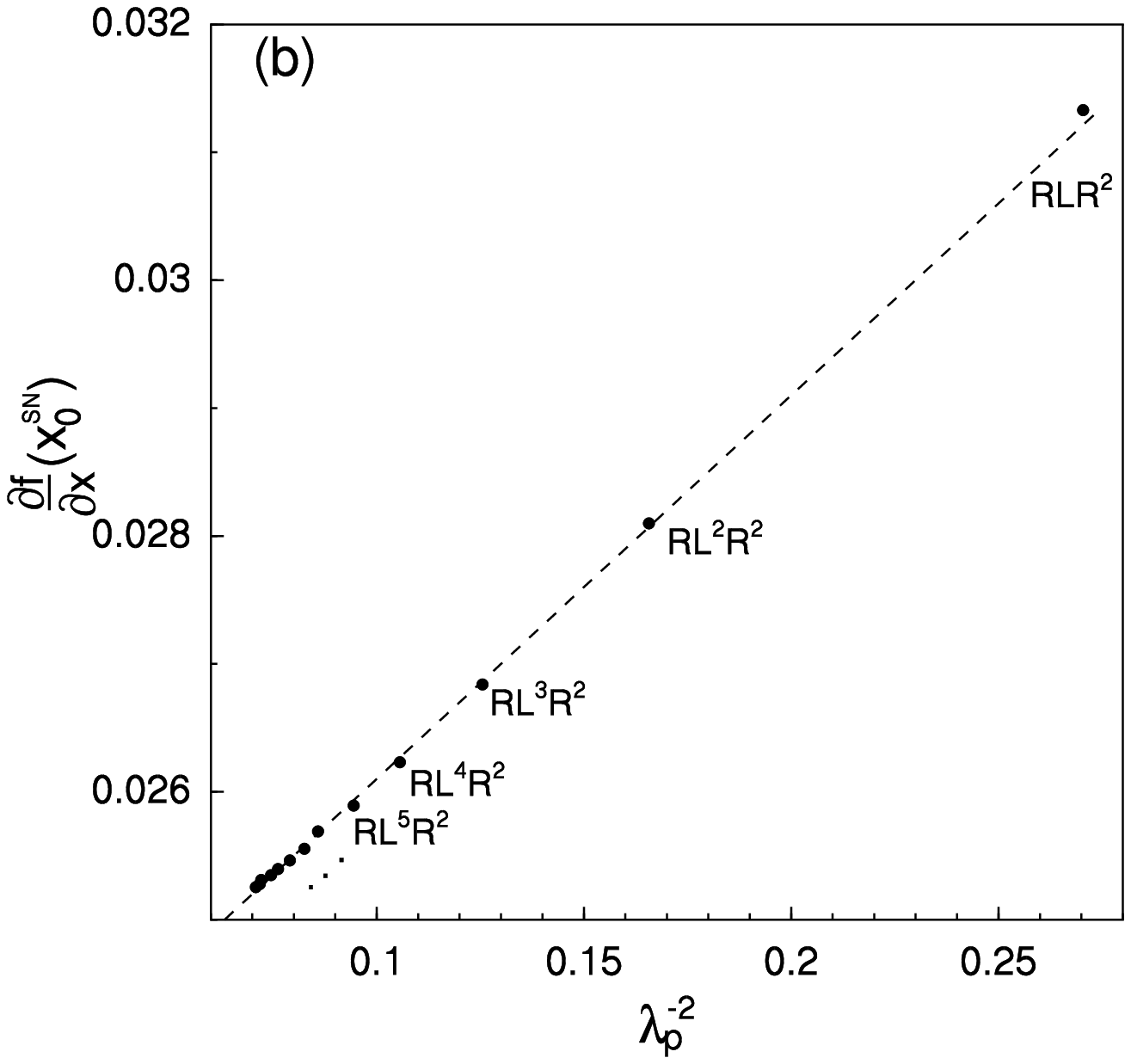}
\caption{(a) Periodic orbit sequence of $RL^4R$ and $RL^4R^2$ for
$\mu=3.990345$ and $\mu=4.001927$ for $\xi=18.0$ for the Map-L (the maps are
indistinguishable from each other on this scale). (b) Verifications of the  scaling relation $\frac{\partial f}{\partial x}(x^{SN}_0)$ with 
$\vert\lambda_p\vert^{-2}$   
for $RL^nR^2$ periodic sequences obtained from Map-L.}
\end{center}
\end{figure}\label{fig8}

It is to be noted that in the chaotic region between
$RL^k$ and $RL^{k+1}$, there exist stable windows of other
symbolic sequences also. In the above, the scaling is shown only
for a particular sequence of periodic windows, namely $RL^kR^m$.
However,  the structure of the map ensures that   any periodic
orbit with periodicity higher than that of  $RL^kR$ and $RL^kR^2$
in the regions $R_I$ and $R_{II}$ respectively, typically have
more number of iterates falling into the neighbourhood $U$ of $p$.
This structural feature along with the large  value of $\lambda_p$
ensures that the other   window widths of these periodic orbits
are increasingly small. //

\section{Concluding Comments}

 One of main motivating factor for the study of a 
specific type of  Poincare maps, namely the NMA maps, of MMO
systems is that these maps constitute the dimensionally reduced
form of the continuous time system; thus these maps  are expected
to retain all the key features of the MMO systems. It is well
known that a dominant feature of homoclinic bifurcation to an
unstable limit cycle (Gavrilov-Shilnikov scenario) \cite{gavrilov}
is the existence of isola structure. This is observed in the
region of periodic-chaotic sequences in MMO systems, and even in
those cases which show incomplete approach to homoclinicity
\cite{petrov92,koper95,raj00}. ( See below.) Since the isola
structure is preserved at the level of long tailed NMA maps,
as  a part of the study of long tailed maps, we have undertaken to
investigate the origin of these isolas by studying the reversal of
bifurcation sequences (within the limited scope of $RL^k$ type of
orbits).
For example, the $RL^k$
sequences were shown to be created and subsequently destroyed
through a fold bifurcation leading to  isola structures. {\it Hence, in
the parameter region where period adding sequences (of type
$RL^k$) manifest, these isolas  are prominent and an isola with
larger periodicity is embedded within the  isola with a lower
periodicity leading to a progressively ordered structure of
isolas.} This feature mimics the ordered structure of the isolas
present in certain types of MMO systems. \\

 From the  point of view of continuous time systems, the
reversal of period doubling    sequences  is a direct consequence
of the back-to-back Hopf bifurcations in these  systems
\cite{koper92,tamosi89,raj00}. For a two parameter dynamical
system,  a degenerate Hopf bifurcation gives rise to back-to-back
Hopf bifurcations  and the existence of the periodic orbits is
confined  to this  interval  in the parameter space.  Within the
region of back-to-back Hopf bifurcations,  any bifurcation wherein
periodic orbits are created  has to be matched by  a reverse
bifurcation of the same kind wherein the  periodic orbits will be
destroyed. In other words, in these continuous flow systems also,
the periodic  orbits born in a fold bifurcation vanishes in
another  fold bifurcation  creating an isola structure. Moreover,
the amplitude of  the period one  orbit arising from the first
Hopf bifurcation approaching the other Hopf bifurcation (while going
from the periodic  state to a  fixed point) gets progressively
smaller. Since the period one orbit is equivalent to a fixed point
of the NMA map, this feature translates to  the condition
$\frac{\partial f}{\partial \mu}(p) < 0$ that has been  imposed
for  the changes in the structure of the map.  A consequence of
this condition  is that {\it the first period doubling occurs  due
to the changes in the  shape of the map in the neighbourhood of
period one fixed point, while the rest of the period doubling
bifurcations are due to the presence of the critical point.}  This
difference in the cause of the first period doubling bifurcation
and   higher order bifurcations is reflected in the first
Feigenbaum   exponent   being distinctly different  from the rest
of the PD bifurcations \cite{rajesht,john}.  Equivalently,  a large period two window
is seen as a   distinct feature  of the alternate
periodic-chaotic sequences of MMO systems also.\\

 We have shown that the increasing sharpness of the map
and the  consequent long tail nature, as the parameter  is tuned,
determines the  dominance of the $RL^k$ periodic states.
Heuristically, the dominance of the $RL^k$ type periodic orbits
can be  related to the arguments on dissipation in the following
manner.  The primary  features of the structure of the map
relevant to the  predominance of period adding sequences is  the
large eigen value (in absolute terms) for  the period one fixed
point, the long tail  structure, and the  presence of an
intermittency channel. In these class of maps, apart from the
neighbourhood of  the critical point, there  exists a large tail
region wherein the slope is  small.  By definition, these regions
are related to  high  dissipation, since the  spread  of the
iterates in its next iteration is  suppressed. (Given an
neighbourhood $I_1$ in the tail region,  $I_2 = f(I_1;\mu,\xi) \ll
I_1$. )  One type  of sequence of periodic orbits which include
visits to the regions of high dissipation with no visits  to the
neighbourhood  containing $p$ are the $RL^k$ sequences. {\it High
dissipation favours periodic orbits, hence the  stability of these
periodic sequences  are enhanced at the expense  of the chaotic
regions.} In the same  spirit, the windows of  periodic orbits
occurring   within the  chaotic region  between any two successive
$RL^k$ sequences,  involving  at least one iterate in  the
neighbourhood of the unstable fixed point and  favouring negative
dissipation (local expansion) have   window widths  smaller by  a
factor of $\lambda_p$. These analytical results  for  the onset of
periodic orbits of the type $RL^kR^m$  show that  the slope
at the point of onset of the periodic orbit, namely
$\frac{\partial f}{\partial x}(x_0)$ is related to the slope of
the  unstable fixed point $\lambda_p$, and scales linearly with
$\lambda_p^{-m}$, where $m$ is the number of visits to the
neighbourhood of the unstable period one fixed point. {\it The
relationship  between  the slopes near the critical point and the fixed point 
essentially
controls the width of such  periodic orbits. It must be
emphasised that these scaling relations derived are very different
from those usually derived where the widths of the periodic orbits
are evaluated as a function of the parameters.} Again, map-L has
been used {\it only} to verify these scaling relations.   The scaling
relation  suggests that  the width of the  parameter window is
nearly independent of the visits to the left of critical  point of
the map  and only the visits to the neighbourhood of  $p$
determine  the width of these type ($RL^k$) of  periodic windows.
For a map with a sharp maximum and a long tail, the  eigen value
of the unstable fixed point will be large  and consequently
smaller windows  widths for the periodic orbits  contained  in the
chaotic region sandwiched between $RL^{k-1}$ and $RL^{k}$.\\

 A necessary condition  for the occurrence of the
reversal of period  doubling sequences  in unimodal maps  have
been  stated in the form of the first two conditions defined by
Eqs. (1) and (2), which are constructed keeping in mind the
changes in the structure of the  NMA maps of the MMO systems.  The
correctness of these conditions  is  validated by the  fact that
the  dominant bifurcation sequences of the long tailed  maps and
the MMO systems have similar features. Further,  
the similarity of the bifurcation diagrams obtained from the
example map-L  ( belonging to the class  of long tailed maps )
is similar to the MMO systems and in particular 
bifurcation sequences from AK model. Since  the dissipation involved in the
higher dimensional continuous time  is related to the separation
of time scales operating in the system, a larger separation of
time scales implies near one dimensional nature  as well as the
long  tailed structure for the map. Since the  scaling relations
incorporate the essential  ingredients of the structure of the
map, namely, the property of long tail and the {\it strength of
the  unstable fixed point} ($\lambda_p$),  the scaling relations
derived for the one-D  maps can be used as an effective check  for
the kind of dynamics involved in the higher dimensional systems.
These observations are subject to usual criticism applicable to
maps considered as reduced dynamical systems.  In higher
dimensional systems, it is conceivable that the interaction
between several parameters can lead to reversal of period doubling
and period adding sequences. The present work only attempts
project the entire complexity into a few constraints. Clearly,
these constraints on the maps can be satisfied
in variety of combinations of parameters in higher dimensional space. 
\\

In summary, we have traced the  dominance of  period
adding sequences as a part of complex alternate periodic-chaotic
sequences in MMO systems to the long tailed nature of the
associated  Poincare maps. In the process, we have {\it partially}
elucidated the origin  of the reversal of the  bifurcation
sequences  with specific reference to $RL^k$ type of orbits which
are  the dominant sequences in MMO systems.  {\it The results are
derived based on  some general  conditions alone without taking
recourse  to any particular analytical form of the  map}; the
map-L is used  only to numerically confirm the analytical results.
Further, the example map-L which satisfies the general constraints
mimics the distinctive  features of higher dimensional continuous
flow  MMO systems, namely,  the reversal of PD bifurcation
sequences and   the alternate periodic-chaotic  sequences. The
specific feature of the dominance of windows periodic orbits over
the chaotic regions in the long tailed  maps  also captures the
equivalent feature for the continuous time MMO systems. Since,
many maps of experimental MMO systems exhibit generic features
similar to those used here, the scaling relations derived for the
onset of  the periodic orbits of the form $RL^kR^m$ can be taken
to serve   as a check for the correctness of the analysis.  In
particular, in BZ systems, while earlier experiments do show  the
dominance of the  period adding sequences \cite{pikovsky}, the
existing  results are not accurate enough to verify these scaling
relations. However,  we believe that careful experiments  should
bear out our results. Indeed, we expect that careful experiments
on  a number of other experimental systems wherein  the
geometrical shape of the Poincare map is similar to Fig. 2
\cite{alba89,koper92,krisher93,othersystems},  could be used to
validate our results. Another example would be the laser systems
where accuracy of control on parameters  is generally considerably
high. These results  also  indicate the ubiquity in  the
qualitative dynamical features of physical systems from widely
differing origin,  exhibiting alternate periodic-chaotic
sequences.

\section{Appendix}
\renewcommand{\theequation}{A-\arabic{equation}}
\setcounter{equation}{0}
In this appendix, we provide an example two parameter map which has the
following broad  typical features observed by the  NMA maps
of MMO systems:
a) non zero origin,
b) a sharp symmetric maximum positioned asymmetrically, and
c) asymptotic long tail   to the right of the maximum.
It is a generalisation of the  Bountis map\cite{bountis} given by
\begin{eqnarray}
           x_{n+1}& = &f(x_n;\mu,\xi)        \hspace{8.0cm} \nonumber          
\\\
                f(x;\mu,\xi)& = &a + \frac{(\xi-\mu)+r(x-c )^l}{s_1 + 
s_2\mu (x-c )^m}  
\end{eqnarray}
\noindent
with $m \ge l$  and $ a > 0$.
We refer to this map given by Eq. (A1) as map-L.   This
map is used mainly to verify the analytical results derived in
text. We shall use only two parameters $\mu$ and $\xi$ of the
several parameters $l,m,a,r,s_1,s_2$ and $c$. The other parameters
of the map $f(x;\mu,\xi)$ are tuned   such that the model  map
looks  similar (in structure) to  one dimensional maps of the MMO
systems (Fig. 2).   For our numerical study, we use the following
values for the map parameters :   $l = m = 4.0, a=3.0, r=0.6,
s_1=1.5, s_2=0.12$ and $c=7.0$. We shall use $\mu$ and $\xi$ as
primary and secondary bifurcation parameters
respectively.\\

 The bifurcation diagrams   are constructed using $\mu$
as a control parameter for fixed value   of $\xi$.  Fig. 3 shows a
typical shape of the map defined by Eq. (A1).     This map has a
unique smooth  maximum  with inflection  points on either side of
the maximum.    For properly chosen initial conditions,   the
asymptotic tail ensures that the  nontrivial dynamics of the map
is restricted to the interval     $[f^2(c;\mu,\xi),
f(c;\mu,\xi)]$.   It can be easily verified that the map has a
negative   Schwarzian derivative everywhere on the positive real
line.  Since $a > 0 $, for an appropriate choice of $a$, there are
no fixed points     in the interval $[0, c]$ for small values of
$\xi$.  For a range of values of the parameter $\xi < \xi^*$,  the
map has only one   fixed point, $p$, which  is in the interval
$[c, f(c)]$.   For  $\xi > \xi^*$, two more fixed points, $p_2$
and $p_3$,   located in the interval $[0,c]$, are   born in a fold
bifurcation at $\mu = \mu^*(\xi)$. ( Both $p$
and $p_3$ are unstable, and $p_2$ is stable.) \\

 In the following, we briefly describe the bifurcation
sequences of the map $f(x;\mu,\xi)$ with respect to the primary
bifurcation parameter   $\mu$ for various values of $\xi$.   For
small but fixed value of $\xi$,   an  increase in the value of
$\mu$ leads to a decrease in the value of   $p$.    For $\xi <
8.0$, the unique fixed point $p$ is stable for all values of
$\mu$. Beyond $\xi = 8.0$, the fixed point $p$ loses stability in
a period doubling bifurcation as $\mu$ is  increased which is
regained in a    period undoubling  bifurcation thus forming a
bubble structure   (Fig. 4a).  For further increase in $\xi$, the
number of nested bubbles grows  as $2^n$ with $n=1,2,3\cdots$. For
$\xi\sim 11.9$, a  period three window   opens up in the
bifurcation diagram separating the period doubling sequences from
the reverse period doubling sequences (Fig. 4b).   Further
increase in the value of $\xi$  unveils  period adding sequences
of arithmetically increasing periodic windows  coexisting with the
bubble structure (Fig. 5a).  This sequence of large parameter
windows of stable  periodic   orbits (in $\mu$)  is found to be of
$RL^k$ type  sequences in the   two letter symbolic dynamics
language. For $\xi = 12.8$,   we find that the  reversal of the
dominant bifurcation sequences occur beyond $\mu \sim 4.05$.   At
$\xi=18.0$ ($\xi^* \sim 13.0$), as we increase $\mu$, at  $\mu =
\mu^* =4.8$,     the region of the map for small values of $x$
makes contact with the bisector  giving rise to a fold bifurcation
with the creation of a pair of stable   and unstable fixed points,
$p_2$ and $p_3$ (Fig. 5b).   Keeping $\xi \ge \xi^*$, as the
parameter $\mu$ is increased towards $\mu^*$, the channel between
the bisector and the map gets progressively  narrow, thereby
stabilising periodic orbits of higher periodicity (See Fig. 3 and
also Figs. 6a,8a).   The positions of the  iterates of  higher
periodic cycles   change marginally to accommodate the next higher
periodic cycle  as  additional iterates are squeezed in this
intermittent channel.    Thus, the intermittency or the fold
bifurcation point ($\mu^*$) is an accumulation point for the
arithmetically increasing period adding sequences. Beyond the
accumulation point,  no further bifurcations are observed  for the
stable periodic point ($p_3$). This map reproduces most bifurcations features of the AK model. ( Compare bifurcation diagrams here with those in \cite{raj00} and \cite{john}.)

\begin{acknowledgements}
One of the authors (R.R) gratefully acknowledge the financial
support from  Department of Science and Technology and partial
financial support from Jawaharlal Nehru Center for Advanced
Scientific Research.
\end{acknowledgements}


\begin{thebibliography}{99}


\bibitem{gyorgi} L. Gyorgi, R. J. Field, Z. Noszticzius, W. D. 
McCormick and H. L. Swinney,
J. Chem. Phys. {\bf 96} 1228 (1992)

\bibitem{barkley} D. Barkley, J. Chem. Phys. {\bf 89} 5547 (1988).

\bibitem{petrov92}
V. Petrov, S.K. Scott,  K. Showalter, J. Chem. Phys.  97, (1992) 6191.

\bibitem{koper95}
M.T.M. Koper, Physica  D80  (1995) 72.

\bibitem{bz}
A. Arneodo, F. Argoul, J. Elezgaray, P. Richetti, Physica D62 (1993) 
134
{\em (and the references therein)}.;
J.S. Turner, J.-C. Roux, W.D. McCormick, H.L. Swinney, Phys. Lett.
  A85  (1981) 9;
P. Ibibson,  S. K. Scott, J. Chem. Soc. Faraday Trans. 87
(1991) 223;
Z. Noszticzius, W.D. McCormick,  H.L. Swinney, J. Phys. Chem. 93
 (1989) 2796.
 

\bibitem{alba89}
F.N. Albahadily, J. Ringland, M. Schell, J. Chem. Phys.
90 (1989) 813.

\bibitem{koper92}
M.T.M. Koper, P. Gaspard,  J. Chem. Phys.  96  (1992) 7797;
M.T.M. Koper, P. Gaspard, J.H. Sluyters, J. Chem. Phys. 97 (1992) 8250.


\bibitem{krisher93}
K. Krischer, M. Lubke, M. Eisworth, W. Wolf, J. L. Hudson, G. Ertl, 
Physica  D62 (1993) 123.


\bibitem{chay} T. R. Chay, Y. S. Yan, and  Y. S. Lee, Int. J. Bifur. 
Chaos {\bf 5} 595 (1995).

\bibitem{braun} T. Braun, J. A. Lisboa, Int. J. Bifur. Chaos {\bf 4} 
1483  (1994).


\bibitem{tamosi89}
Ferdinando de Tamosi, D. Hennequin, B. Zambon, E. Arimondo, J. Opt. 
Soc. Am.  B6
(1989) 45.

\bibitem{raj00}
S. Rajesh and G. Ananthakrishna, Physica D140 (2000) 193;
S. Rajesh and G. Ananthakrishna, Phys. Rev. E  61 (2000) 3664.


\bibitem{hauser96}
M.J.B. Hauser, L.F. Olsen, J. Chem. Soc., Faraday Trans.
 92 (1996) 2857.


\bibitem{rajesh99}
S. Rajesh,  G. Ananthakrishna, Physica  A270 (1999) 182.

\bibitem{ana82}
G. Ananthakrishna,  M.C.  Valsakumar,  J. Phys.  D15 (1982) L171.
The basic model was formulated in G. Ananthakrishna, D. Sahoo, J. Phys.
 D14 (1981) 2081.


\bibitem{lorenz63}
L.N. Lorenz, J.Atmos. Sci. 20 (1963) 130.


\bibitem{rajesht} S. Rajesh, Ph. D Thesis, Indian Institute of
Science, Bangalore (2000).

\bibitem{coffman86}
K.G. Coffman, W.D. McCormick, H.L. Swinney, Phys. Rev. Lett.  56 (1986) 
999;
K.G. Coffman, W.D. McCormick, Z. Noszticzius, R. H. Simoyi, H.L. 
Swinney, J. Chem. Phys. 86 (1987) 119.


\bibitem{othersystems}
M.R. Bassett,  J.L. Hudson, J. Phys. Chem.  92  (1988) 6963;
Y. Xu,  M. Schell, J. Phys. Chem.  94  (1990) 7137;
F. Argoul, J. Huth, P. Merzeau, A. Arneodo,  H.L. Swinney, Physica  D62 
(1993) 170;
T. Braun, J.A. Lisboa, Int. J. Bif. and Chaos 4  (1994) 1483;
C.J. Doona,  S.I. Doumbouya, J. Phys. Chem.  98, (1994) 513.




\bibitem{guc90}
J. Guckenheimer, P.J. Holmes, Nonlinear oscillations, Dynamical
Systems and Bifurcations of Vector Fields, Springer, Berlin, 1990.



\bibitem{gaspard84}
P. Gaspard, R. Kapral, G. Nicolis, J. Stat. Phys. 35 (1984) 697.

\bibitem{glenn84}
P. Glenndining, C. Sparrow, J. Stat. Phys. 35 (1984) 645.


\bibitem{gaspard87}
P. Gaspard, X.J. Wang, J. Stat. Phys. 48 (1987) 151.

\bibitem{michelin94}
O. Michelin, P.E. Phillipson, Phys. Lett.  A188 (1994) 309.

\bibitem{fang95}
H.P. Fang, Z. Phys. B 96 (1995) 547.

\bibitem{simoyi82}
R.H. Simoyi, A. Wolf, H.L. Swinney, Phys. Rev. Lett.  49  (1982) 245.

\bibitem{bagley86}
R.J. Bagley, G. Mayer-Kress, J.D. Farmer, Phys. Lett. A114  (1986) 419.

\bibitem{ring84}
J. Ringland, J.S. Turner, Phys. Lett. A105 (1984) 93.

\bibitem{pikovsky}
A.S. Pikovsky, Phys. Lett.  A85 (1981) 13.

\bibitem{bountis}
M. Bier,  T.C. Bountis, Phys. Lett.  A104  (1984) 239.

\bibitem{nusse}
H.E. Nusse, J.A. Yorke, Phys. Lett.  A127 (1988) 328.

\bibitem{parlitz}
U. Parlitz, W. Lauterborn, Phys. Lett.  A107 (1985) 351; Phys. Rev.
 A36 (1987) 1428.

\bibitem{kan92}
I. Kan, H. Kocak, J.A. Yorke, Ann. Math. 136  (1992) 219.


\bibitem{dawson}
S.P. Dawson, C. Grebogi, J.A. Yorke, I. Kan,  H. Kocak, Phys.
Lett.  A162 (1992) 249;
S.P. Dawson, C. Grebogi, H. Kocak, Phys. Rev.  E48  (1993) 1676.

\bibitem{ringland}
J. Ringland, N.Issa, M. Schell, Phys. Rev.  A41 (1990) 4223.


\bibitem{schi88}
M. Schintuchand and Schimdt, J. Phys. Chem. {\bf 92} 3404 (1988).

\bibitem{geisel81}
T. Geisel, J. Nierwetberg, Phys. Rev. Lett.  47 (1981) 975.

\bibitem{mss}
N. Metropolis, M.L. Stein,  P.R. Stein, J. Comb. Theory
15  (1973) 25.

\bibitem{post91}
T. Post, H.W. Capel, Physica  A178 (1991) 62.

\bibitem{gardini}
L. Gardini, Nonlinear Anal., Theory, Methods and App. 23 (1994) 1039.

\bibitem{gavrilov}
N.K. Gavrilov, L.P. Shilnikov, Mat. USSR Sb. 17
(1972) 467; 19 (1973) 139.

\bibitem{aguda91}
B.D. Aguda, R. Larter, J. Am. Chem. Soc.  113 (1991) 7913.

\bibitem{newell96}
T.C. Newell, V. Kovanis, A. Gavrielides,  P. Bennet, Phys. Rev.  E54  
(1996) 3581.

\bibitem{tomita80}
K. Tomita, I. Tsuda, Prog. Theo. Physics  64  (1980) 1138.

\bibitem{circlemap}
K. Kaneko, Prog. Theor. Phys.  68 (1982) 669; K. Kaneko, Prog.
Theor. Phys. 69 (1983) 403.

\bibitem{corbet}
A.B. Corbet, Phys.  Lett. A130 (1988) 267.

\bibitem{swinney}
H.L. Swinney, Physica  D7  (1983) 3; K. Coffman, W. D. McCormick,
H.L. Swinney, Phys. Rev. Lett. 56 (1986) 999.

\bibitem{john}
G. Ananthakrishna, T. M. John, in: Hao Bai-Lin (Ed.), Directions in chaos, World Scientific, Singapore, 1990.

\end{thebibliography}
\end{document}